\documentclass[preprint]{imsart}

\usepackage{amssymb}
\usepackage{amsthm,amsmath,natbib}
\RequirePackage[colorlinks,citecolor=blue,urlcolor=blue]{hyperref}
\usepackage{graphicx}
\usepackage[colorinlistoftodos]{todonotes}
\usepackage{eucal} 
\usepackage{outlines}
\usepackage[caption=false]{subfig}
\usepackage{algorithm}
\usepackage{algorithmic}
\usepackage{mathtools}
\usepackage{booktabs}
\usepackage{multirow}
\usepackage{ragged2e}
\usepackage{ulem}
\usepackage{color}
\usepackage{rotating}
\usepackage{xr}
\usepackage{color,soul}

\graphicspath{ {images/} }

\startlocaldefs
\newcommand{\A}{\ensuremath{\mathtt{A}}}
\newcommand{\C}{\ensuremath{\mathtt{C}}}
\newcommand{\G}{\ensuremath{\mathtt{G}}}
\newcommand{\T}{\ensuremath{\mathtt{T}}}

\newcommand{\N}{\ensuremath{\mathtt{N}}}
\newcommand{\R}{\ensuremath{\mathtt{R}}}
\newcommand{\Snuc}{\ensuremath{\mathtt{S}}}
\newcommand{\W}{\ensuremath{\mathtt{W}}}
\newcommand{\Y}{\ensuremath{\mathtt{Y}}}
\newcommand{\MK}{\ensuremath{\mathtt{MK}}}

\newcommand{\M}{\ensuremath{\mathcal{M}}}

\endlocaldefs

\begin{document}

\begin{frontmatter}

\title{Survival analysis of DNA mutation motifs with penalized proportional hazards}
\runtitle{Survival analysis of mutation motifs}

\author{\fnms{Jean} \snm{Feng}${}^\S$},
\address{Department of Biostatistics, University of Washington\\ Seattle, WA, USA}
\author{\fnms{David A.} \snm{Shaw}${}^\S$},
\address{Computational Biology Program, Fred Hutchinson Cancer Research Center\\ Seattle, WA, USA}
\author{\fnms{Vladimir N.} \snm{Minin}*\ead[label=e2]{vminin@uci.edu}},
\address{Department of Statistics, University of California, Irvine\\ Irvine, CA, USA\\ \printead{e2}}
\author{\fnms{Noah} \snm{Simon}*\ead[label=e1]{nrsimon@u.washington.edu}},
\address{Department of Biostatistics, University of Washington\\ Seattle, WA, USA\\ \printead{e1}}
\vspace{0.2cm} \and \vspace{-0.15cm}
\author{\fnms{Frederick A.} \snm{Matsen IV}*\ead[label=e3]{matsen@fredhutch.org }},
\address{Computational Biology Program, Fred Hutchinson Cancer Research Center\\ Seattle, WA, USA\\ \printead{e3}}

\runauthor{Feng and Shaw et al.}

\vspace{0.3cm}

\begin{centering}
\text{\S \ Co-first authors} \\
\text{* \ Co-corresponding authors}
\end{centering}

\begin{abstract}
Antibodies, an essential part of our immune system, develop through an intricate process to bind a wide array of pathogens.
This process involves randomly mutating DNA sequences encoding these antibodies to find variants with improved binding, though mutations are not distributed uniformly across sequence sites.
Immunologists observe this nonuniformity to be consistent with ``mutation motifs'', which are short DNA subsequences that affect how likely a given site is to experience a mutation.
Quantifying the effect of motifs on mutation rates is challenging: a large number of possible motifs makes this statistical problem high dimensional, while the unobserved history of the mutation process leads to a nontrivial missing data problem.
We introduce an $\ell_1$-penalized proportional hazards model to infer mutation motifs and their effects.
In order to estimate model parameters, our method uses a Monte Carlo EM algorithm to marginalize over the unknown ordering of mutations.
We show that our method performs better on simulated data compared to current methods and leads to more parsimonious models.
The application of proportional hazards to mutation processes is, to our knowledge, novel and formalizes the current methods in a statistical framework that can be easily extended to analyze the effect of other biological features on mutation rates.
\end{abstract}

\end{frontmatter}

\section{Introduction}

We introduce a proportional hazards model approach to study DNA mutation processes.
Our study is motivated by somatic hypermutation, a mutation process that occurs in DNA sequences that encode B-cell receptors (BCRs), proteins that recognize and neutralize pathogens.
When BCRs are secreted from B cells they are known as antibodies.
The immune system relies on this somatic hypermutation process to generate a diversity of BCRs that can bind to a large and continually evolving variety of pathogens.
The starting material for this mutation process is a BCR sequence that is formed by recombination \citep{Tonegawa1983-qz,Schatz2011-cm}.
From this sequence, a complex system of enzymes introduces mutations in a random pattern that is known to be highly sensitive to the sequence \textit{motif}: the sequence of DNA bases surrounding the mutating position \citep{Dunn-Walters1998-ds,Chahwan2012-gp, Rogozin1992-xv,Methot2017-gi}.

Our goal is to develop a solid statistical framework that estimates the mutation rates of motifs and provides interpretable results for this mutation process.
A better understanding of somatic hypermutation will help in designing vaccines for challenging viruses \citep{Haynes2012-zo,Hwang2017-tt,Wiehe2018-of}, in furthering understanding of the biological mechanisms at play \citep{Pham2003-jm, Rogozin2001-pw}, and in gaining insight into the natural selection process occurring in the immune system \citep{Hershberg2008-rp,Uduman2011-ib,McCoy2015-qi,Hoehn2017-ol}.

Several strategies have been used to model a motif's mutability -- that is, how likely a position is to mutate given the motif at that position.
The general approach is to compare a mutated sequence with its inferred ancestor sequence and model the differences between them.
\citet{Cohen2011-rs} and \citet{Elhanati2015-ld} model the mutabilities of motifs as the product of the mutabilities of short subsequences (usually 1 or 2 bases).
By using a log-linear model with only first-order terms they keep the parameter count low, but miss interactions between the positions.
\citet{Yaari2013-dg} and \citet{Cui2016-wz} do not use this log-linear assumption: they allow a separate parameter for each possible five-nucleotide motif (of which there are $4^5$), and use ad-hoc methods to handle motifs with few observations.
Rather than these restrictive and ad-hoc approaches, a more data-adaptive variable selection method is desirable.

Another drawback of these methods is that they ignore mutations that occur in neighboring positions, even though such events can carry important information about highly mutable motifs.
Indeed, these methods require counting the number of times a motif is observed to mutate: if mutations occur in neighboring positions, they cannot attribute the mutation to the correct motif.
For settings with high rates of mutation, these methods end up estimating the mutabilities poorly.
To properly estimate mutabilities, one needs to account for the different possible orders that mutations occurred in.
Previous work has developed methods for performing various types of inference when this mutation order is unknown \citep{Hwang2004-pj, Hobolth2008-dx}, but these inference procedures make the parametric assumption that the mutation process follows a continuous time Markov process.
Here we relax this model assumption and use a semiparametric model instead.

In this paper, we advance the modeling of motif mutabilities in several directions.
We propose a method to fit mutabilities using \underline{s}urvival \underline{a}nalysis of \underline{m}utation \underline{m}otifs, called \texttt{samm}.
We formalize the problem using Cox proportional hazards, in which mutations are the failure events to be investigated.
Although survival models are used implicitly by computational immunologists for simulation \citep{Yaari2012-kk,Sheng2017-ib}, we believe this is the first time they have been used for inference.

To estimate motif mutabilities, our method uses the Monte Carlo expectation--maximization algorithm \citep[MCEM,][]{wei1990monte}.
Since the orders in which mutations occur are unobserved in our data, expectation--maximization \citep[EM,][]{dempster1977maximum} allows us to perform maximum likelihood while averaging over these unknown orders.
However the E-step in EM requires calculating the expected log-likelihood, which is analytically intractable since we must average over all possible mutation orders; thus we estimate this expectation using Gibbs sampling.
This approach is similar to that used by \citet{goggins1998markov} to model interval-censored failure-time data where the order in which the failure events occur is unknown.

Our method also handles high-dimensional settings in which there are many more predictors than observations, which is important because many motifs are hypothesized to affect the mutation rate but the specific ones are unknown.
For instance, \citet{Yaari2013-dg} and \citet{Cui2016-wz} consider all motifs of length 5.
We use the lasso \citep{tibshirani1996regression} to improve estimation and perform variable selection.
To provide a measure of uncertainty of our estimates, we use a two-step approach: we fit an $\ell_1$-penalized Cox proportional hazards model \citep{tibshirani1997lasso} to perform variable selection and refit an unpenalized model over the selected variables to obtain our final estimates along with approximate confidence intervals.

Section~\ref{sec:methods} describes our estimation methods, starting with a simplified logistic regression model and then progressing to our full estimation method.
Section~\ref{sec:simulations} presents simulation results.
In Section~\ref{sec:data}, we apply our method to model somatic hypermutation of BCR sequences from \citet{Cui2016-wz} and compare results with previous methods.

\section{Methods}
\label{sec:methods}

Our data consists of BCR nucleotide sequences that have mutated for an unknown period of time.
Specifically, we target sequences that are undergoing mutation but not natural selection.
Such data can be obtained, for example, through immunization experiments in transgenic mice designed to have a DNA segment that is carried along and mutated but not expressed as part of the BCR \citep{Yeap2015-nl,Cui2016-wz}.

Though we focus on modeling the somatic hypermutation process of BCRs, our approach can be framed more generally as a problem of modeling a sequence-valued mutation process.
We refer to the original, unmutated sequences as ``na\"ive'' and their descendants as ``mutated'' sequences.
Throughout, we suppose that these na\"ive sequences are known.
In the BCR case, we restrict our attention to a computationally-identified na\"ive segment coded in germline DNA \citep[the V region,][]{Yaari2015-pc}.

More formally, the mutation process of a sequence with $p$ positions can be described as a vector-valued stochastic process $\{X(t) = (X_1(t), ..., X_p(t)): t \in [0,\infty)\}$ indexed by time $t$.
Each $\{X_j(t)\}$ represents the mutation process of the $j$th position in the sequence.
For a given time $t$, the state space of $X_j(t)$ is the set of nucleotides $\{\A, \C, \G, \T\}$ and the state space of $X(t)$ is the set of length-$p$ nucleotide sequences $\mathcal{S} = \{\A, \C, \G, \T\}^p$.
At the start of the mutation process, $X(0)$ is fixed to be the na\"ive sequence.

In a context-sensitive model, the probability that a position mutates at time $t$ depends on the current nucleotide sequence $X(t)$.
In our work, we assume that only local context matters: The mutation rate at each position is affected only by the local nucleotide sequence called the motif. For motif $m$, we denote the length of the motif as $\operatorname{len}(m)$, where $\operatorname{len}(m)$ is typically much smaller than the number of nucleotides in $X(t)$.
The function $I(X(t), m, j, j')$ is the binary indicator of whether motif $m$ appears in sequence $X(t)$ from positions $j - j' + 1$ to $j  - j' + \operatorname{len}(m)$.
More formally, it is defined as
\begin{align}
I(X(t), m, j, j') = \prod_{k=1}^{\operatorname{len}(m)} 1 \left \{X_{j- j' + k}(t) = m_{k} \right \},
\label{eq:indicator}
\end{align}
where $m_{k}$ is the nucleotide in the $k$th position of motif $m$.
This is known as a $\operatorname{len}(m)$-mer, i.e., a motif of length $\operatorname{len}(m)$.
For example, a 5-mer is a motif of length 5.
In the special case where $\operatorname{len}(m)$ is odd and $j' = (\operatorname{len}(m)+1) / 2$, \eqref{eq:indicator} checks if $X(t)$ has motif $m$ centered at position $j$.
We call this a centered motif; for all other cases we say that \eqref{eq:indicator} is checking for an offset motif.

\begin{figure}
	\includegraphics[width=0.6\textwidth]{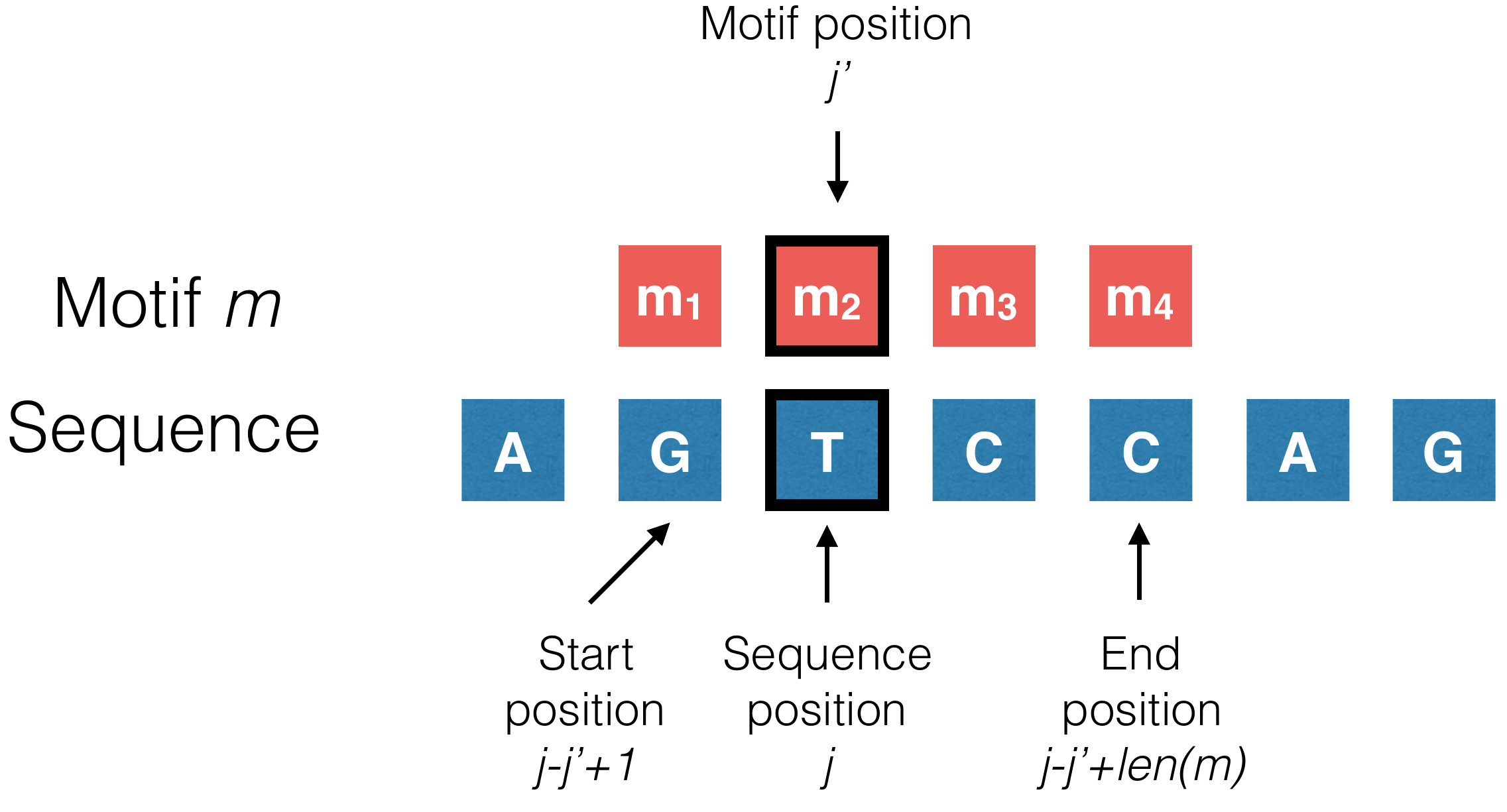}
	\caption{
		An example of how feature vectors are generated: if we believe that the mutation rate at a position depends on the 4-mer (i.e.\ length 4 motif) starting one position to its left, then the feature vector for position $j$ is a one-hot encoding of the sequence that appears in position $j - 1$ through $j + 2$.
		More formally, each element in the feature vector at position $j$ indicates whether or not a motif $m$ appears from start position $j - j' + 1$ through end position $j - j' + \operatorname{len}(m)$ (here $m = 4$ and $j' = 2$).
		The start and end positions are derived by aligning position $j$ of the sequence with position $j'$ of the motif.
	}
	\label{fig:indicator}
\end{figure}

Define a \textit{motif dictionary} to be a set $\M$ of sequence features $(m, j')$ that may affect mutation rate.
Example dictionaries include $1$-mers (all length $1$ motifs), offset $2$-mers (length $2$ motifs with $j'=1,2$), all of the central and offset $3$-mers (length $3$ motifs, with $j' = 1,2,3$), and all of the central $5$-mers.
We may also consider all possible unions of these dictionaries.
Suppose we have selected a set $\M$.
To ease exposition, we choose an arbitrarily assigned but fixed order $\left\{(m^{(1)},j'^{(1)}),\ldots, (m^{(q)}, j'^{(q)})\right\}$ where $q$ is the number of motif features in the dictionary $\M$.
We may now define a function that indicates which elements in $\M$ occur at each position.
For each position $j$, let $\psi_j: \mathcal{S} \mapsto \left\{0,1\right\}^q$ be defined by $\left[\psi_j(X(t))\right]_k \equiv I(X(t), m^{(k)}, j, j'^{(k)})$ for $k = 1,...,q$.
We use $\psi_j$ as the feature vector for modeling the mutation rate of position $j$ (Figure~\ref{fig:indicator}).

Of course, the framework we present here generalizes to other types of dictionaries, including dictionaries that only specify bases for a subset of positions, but we will restrict to the above-described dictionaries in this paper for concreteness.

\subsection{Logistic regression}
\label{sec:logistic}

As a simplified approach to modeling the mutation process, one may ignore the time component and use logistic regression.
In this model, each position in the sequence is independent and the probability of each position mutating only depends on the \textit{initial} nucleotide sequence $X(0)$, i.e.
\begin{align}
\Pr(\text{mutation at position } j) = \frac{1}{1 + \exp(-\boldsymbol{\theta}^\top \psi_j(X(0)))} \qquad \forall j \in \{1,...,p\}.
\label{eq:logistic}
\end{align}
\citet{Yaari2013-dg} essentially take this approach; the logistic model here just formalizes their intuition within a statistical framework and allows us to generalize their method to be applicable for any feature vector mapping.
Moreover, we can use penalized logistic regression for handling high-dimensional models and encode various structural assumptions regarding the mutation rates; we discuss this in detail later in Section~\ref{sec:lasso}.

Logistic regression ignores the time component in a mutation process, and as such ignores how the mutation rate of each position may change as other positions mutate (Figure~\ref{fig:surv_anal}).
The assumption that the mutation rate only depends on the initial nucleotide sequence is most problematic when the mutation rate is high.
Also, logistic regression ignores censoring: the method estimates the average mutation probability with respect to a particular sampling process.
The estimates will be different if we tend to sample sequences that mutate for long vs.\ short periods of time.
The following section addresses these issues by modeling the mutation process using a survival analysis framework.

\subsection{Cox proportional hazards}\label{sec:cox}

\begin{figure}
    \centering
    \includegraphics[width=0.5\textwidth]{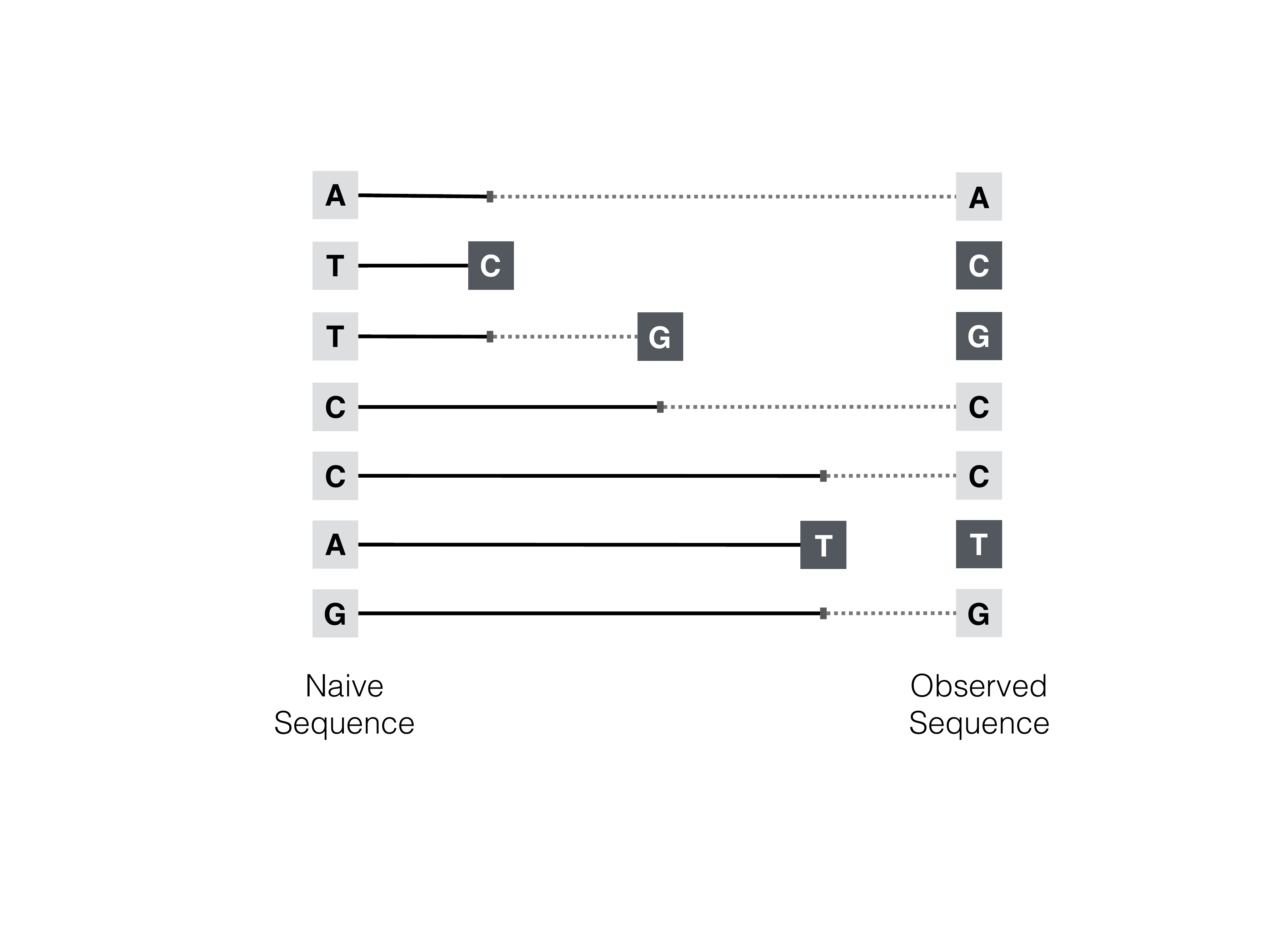}
    \caption{
        Survival analysis for BCR sequences, where the positions that have not mutated are indicated by light gray squares and those that have mutated are indicated by dark gray squares.
        In a context-dependent mutation model, a mutation event can change the mutation rates of other positions.
        Suppose the hazard (i.e.\ mutation) rate of a position depends on the position's two neighboring bases.
        Then, for example, when the \T\ in the third position mutates to a \G, the hazard rate for \C\ in the fourth position changes from the original $\mathtt{TCC}$ motif to the $\mathtt{GCC}$ motif.
        Changes in the motif at a potential mutating position, and thus its hazard rate, are indicated by a change from solid to dashed lines.
    }
    \label{fig:surv_anal}
\end{figure}

We propose using a survival analysis framework to model the mutation process.
We view positions in a single sequence as subjects observed for the same time period.
A mutation event at position $j$ occurs at time $t$ if the nucleotide immediately before time $t$, $\lim_{s \rightarrow t^-} X_j(s)$, differs from the nucleotide at time $t$, $X_j(t)$.
If a position never mutates, we consider its mutation time to be censored.%
\footnote{
By using a survival analysis framework, we implicitly assume that a mutation will occur at every position given a sufficiently long period of time.
This assumption is reasonable for somatic hypermutation -- the complex system of enzymes has the ability to mutate any position along the sequence \citep{Chahwan2012-gp}.
This assumption may not hold for other DNA mutation processes, and the method may need to be modified accordingly.
}
The hazard (or mutation) rate of a position is the instantaneous risk of mutating at time $t$ given that it has been conserved up to time $t$.
In between successive mutation times, each position has a constant hazard rate, and mutates independently from all other positions.
The dependence between positions is introduced when a mutation occurs: upon a mutation event, the hazard rate for each neighboring position can change (Figure~\ref{fig:surv_anal}).

Accounting for how the sequence can change over time complicates our estimation procedure.
Since we do not observe the order of mutation events in the data -- we only observe pairs of na\"ive and mutated sequences -- there are many possible mutation orders that could explain how the mutated sequence arose from the na\"ive sequence;  each mutation order corresponds to a distinct sequence of hazard rates.

For ease of exposition, we present our estimation method for the mutation process of a single pair of na\"ive and mutated nucleotide sequences.
The method readily applies to estimating rates given many independent mutation processes (a typical application will be to thousands or more sequences).

As part of our modeling framework, we assume that each position can mutate at most once during the mutation process.
This is a simplification of the somatic hypermutation process since it is possible for a position to mutate more than once, though in our data the na\"ive and mutated sequence typically differ in 1--5\% of the positions.
We think this assumption is reasonable and makes the problem easier to handle from a computational standpoint.
We discuss how this assumption affects performance under model misspecification in Section~\ref{sec:mutating_replacement} of the Appendix.

We model the hazard rate of position $j$ using Cox proportional hazards, which supposes that the hazard rate $j$ at time $t$ is assumed to be of the form
\begin{equation}
h_{j}(t)=h_{0}(t)\exp\left(\boldsymbol{\theta}^{\top}\psi_{j}\left(X(t)\right)\right),
\label{eq:hazard_rate}
\end{equation}
where $\boldsymbol{\theta}\in\mathbb{R}^q$ and the baseline hazard rate $h_{0}(\cdot)$ is an arbitrary unspecified baseline hazard function.
Extending \eqref{eq:hazard_rate}, we can additionally model the rate at which our process mutates to a specific nucleotide -- the \textit{target nucleotide}.
Previous work \citep{Cowell2000-rq,Rogozin2001-pw,Yaari2013-dg,Cui2016-wz} suggests that the context-dependent mutation process is biased in favor of mutations to particular bases.
We can take into account these preferences by considering a \textit{per-target model}.
In such a model, we additionally define vectors $\boldsymbol{\theta}_{\N}$ for each possible target nucleotide $\N\in\{\A, \C, \G, \T\}$.
Using a competing events framework, the rate of mutating to nucleotide $\N$ at position $j$ at time $t$ is modeled as
\begin{align}
h_{j, \rightarrow \N}(t) = {1}\{X_j(t) \ne \N \} h_0(t) \exp\left(\left(\boldsymbol{\theta} + \boldsymbol{\theta}_{\N} \right )^\top \psi_{j}(X(t))\right).
\label{eq:hazard_rate_target}
\end{align}
As $\N\rightarrow\N$ is not considered a mutation, we include the indicator function $1\{\cdot\}$ in \eqref{eq:hazard_rate_target} to specify that a position cannot mutate to the nucleotide that currently appears there.

\subsection{Maximum likelihood via MCEM}

We are now ready to present a maximum likelihood estimation method for our model.
We assume that the hazard rate follows \eqref{eq:hazard_rate}.
The per-target model in \eqref{eq:hazard_rate_target} is a straightforward extension of this simpler case.
Let the observed data, namely the single pair of na\"ive and mutated nucleotide sequences, be denoted $\mathbf{S}_{\text{obs}}$, where we suppose that $n$ positions have mutated.

When $h_0(t)$ is an arbitrary unspecified baseline hazard function, only the order of the mutations carries information about $\boldsymbol{\theta}$, even if the mutation times are observed \citep{kalbfleisch2011statistical}.
Explained intuitively, time can be transformed by an arbitrary increasing function and the form of the hazard function would still be of the form \eqref{eq:hazard_rate}.
\citep[For more details, see Chapter 4 in][]{kalbfleisch2011statistical}.
Consequently, estimating $\boldsymbol{\theta}$ involves only maximizing the likelihood of observing the mutation order.

For now, suppose we observe the order that the mutations occurred in.
Let $\pi_j$ be the position of the $j$th mutation for $j = 1,...,n$.
Let $\boldsymbol{\pi}_{1:j}$ denote the positions of the first through $j$th mutation, where $\boldsymbol{\pi}_{1:0}$ is defined to be the empty set.
Define $S(\boldsymbol{\pi}_{1:j})$ to be the nucleotide sequence after positions $\boldsymbol{\pi}_{1:j}$ mutate.
Thus the observed data is $\mathbf{S}_{\text{obs}} = \{S(\pi_{1:0}), S(\pi_{1:n})\}$.
The set $R(\boldsymbol{\pi}_{1:j}) \equiv \{1,...,p\}\setminus \boldsymbol{\pi}_{1:j} $ is the set of positions at risk of mutating, commonly referred to as the risk group in the survival analysis literature.
Then the marginal likelihood of $\boldsymbol{\theta}$ is
\begin{align}
\label{eq:likelihood_complete_data}
\mathcal{L}_{c}(\mathbf{S}_{\text{obs}} , \boldsymbol{\pi}; \boldsymbol{\theta})
=
\prod_{j=1}^n
\frac{
	\exp({\boldsymbol{\theta}}^\top \psi_{\pi_{j}}(S(\boldsymbol{\pi}_{1:j - 1})))
}{
	\sum_{k \in R(\boldsymbol{\pi}_{1:j- 1})}
	\exp({\boldsymbol{\theta}}^\top \psi_{k}(S(\boldsymbol{\pi}_{1:j- 1})))
}.
\end{align}
Our result looks like the marginal likelihood derived in equation 4.47 in \citep{kalbfleisch2011statistical} except that it is derived under a more general set of assumptions -- whereas they assume the covariates are fixed, we assume the covariates to be fixed between events.
The derivation of \eqref{eq:likelihood_complete_data} is given in the Appendix.

The marginal likelihood in \eqref{eq:likelihood_complete_data} implies that the mutation order can be simulated by drawing positions from successive multinomial distributions.
To simulate mutation at the $j$th position, we draw a position from the risk group $R(\boldsymbol{\pi}_{1:j-1})$.
In fact, \citet{gupta2015change} use this procedure to simulate the somatic hypermutation process, though they do not provide a statistical justification.

Unfortunately the mutation order $\boldsymbol{\pi}$ is not observed in our problem.
We instead maximize the observed data likelihood, which is the complete data likelihood marginalized over all admissible mutation orders $\mathcal{A}(\mathbf{S}_{\text{obs}})$:
\begin{align}
\label{eq:likelihood_observed_data}
\mathcal{L}(\mathbf{S}_{\text{obs}}; \boldsymbol{\theta})
=
\sum_{\boldsymbol{\pi} \in \mathcal{A}(\mathbf{S}_{\text{obs}})}
\mathcal{L}_{c}(\mathbf{S}_{\text{obs}} , \boldsymbol{\pi}; \boldsymbol{\theta})
.
\end{align}
Assuming positions mutate at most once, $\mathcal{A}(\mathbf{S}_{\text{obs}})$ is a set of $n!$ possible mutation orders.
When the number of mutated positions $n$ is small, we can enumerate all possible mutation orders and maximize \eqref{eq:likelihood_observed_data} using a nonlinear optimization algorithm such as EM \citep{dempster1977maximum}.
However, in most data sets, $n$ is much too large for direct enumeration to be computationally tractable, so we maximize \eqref{eq:likelihood_observed_data} using MCEM.

MCEM extends the traditional EM algorithm by approximating the expectation in the E-step using a Monte Carlo sampling method.
Let $\boldsymbol{\pi} = \boldsymbol{\pi}_{1:n}$ be a full mutation order.
We use the Gibbs sampler in Algorithm~\ref{algo:gibbs} to sample $\boldsymbol{\pi}\mid\{\mathbf{S}_{\text{obs}},\boldsymbol{\theta}\}$.
Given a full mutation order $\boldsymbol{\pi}$, let $\boldsymbol{\pi}_{(-j)}$ be the partial mutation order where the $j$th mutation is removed from $\boldsymbol{\pi}$; a full mutation order $\boldsymbol{\pi}'$ is consistent with $\boldsymbol{\pi}_{(-j)}$ if there is some $j' \in \{1,...,n\}$ such that $\boldsymbol{\pi}'_{(-j')} = \boldsymbol{\pi}_{(-j)}$.
For instance, if $\boldsymbol{\pi} = [1,3, 2]$ then the partial mutation order $\boldsymbol{\pi}_{(-2)}$ is $ [1,2]$ and $\boldsymbol{\pi}' = [3,1,2]$ is consistent with $\boldsymbol{\pi}_{(-2)}$ since $\boldsymbol{\pi}_{(-2)}=\boldsymbol{\pi}'_{(-1)}$.
For each Gibbs sweep, the index $j$ cycles through $\{1,...,n\}$ in some random order.
For Gibbs step $k$, we sample a full mutation order $\boldsymbol{\pi}^{(k)}$ that is consistent with the partial mutation order $\boldsymbol{\pi}^{(k - 1)}_{(-j)}$.
The proof that this sampler converges to the desired probability distribution is standard and similar to that of \citet{goggins1998markov}.

We efficiently calculate the probability of a full mutation order given a partial mutation order by reusing previous computations.
In particular, for partial mutation order $\boldsymbol{\pi}_{(-j)}$, we calculate the probabilities of each consistent full mutation order starting from the full mutation order where position $\pi_j$ mutates first to that where position $\pi_j$ mutates last.
By ordering consistent full mutation orders in this way, the $j'$th consistent full mutation order $\boldsymbol{\pi}'$ and $(j' + 1)$th consistent full mutation order $\boldsymbol{\pi}''$ are the same except that the $j'$ and $(j' + 1)$th mutations are swapped.
The ratio of the conditional probabilities of $\boldsymbol{\pi}'$ and $\boldsymbol{\pi}''$ given $\boldsymbol{\pi}_{(-j)}$ is
\begin{equation}
\begin{split}
\frac{
\Pr(\boldsymbol{\pi}' | \boldsymbol{\pi}_{(-j)})
}{
\Pr(\boldsymbol{\pi}'' | \boldsymbol{\pi}_{(-j)})
}
& =
\frac{
\exp\left (
\boldsymbol{\theta}^\top
\left(
\psi_{\pi_{j'}'}\left(S(\boldsymbol{\pi}_{1:j' - 1}')\right)
+ \psi_{\pi_{j' + 1}'}\left(S(\boldsymbol{\pi}_{1:j'}')\right)
\right)
\right)
}{
\exp\left (
\boldsymbol{\theta}^\top
\left(
\psi_{\pi_{j'}''}\left(S(\boldsymbol{\pi}_{1:j' - 1}'')\right)
+ \psi_{\pi_{j' + 1}''}\left(S(\boldsymbol{\pi}_{1:j'}'')\right)
\right)
\right)
} \times\\
&\qquad \qquad \qquad \frac{
\sum_{i \in R(\boldsymbol{\pi}''_{1:j'})}
\exp\left (
\boldsymbol{\theta}^\top
\psi_{i}\left(S(\boldsymbol{\pi}''_{1:j'})\right)
\right)
}{
\sum_{i \in R(\boldsymbol{\pi}_{1:j'}')}
\exp\left (
\boldsymbol{\theta}^\top
\psi_{i}\left(S(\boldsymbol{\pi}'_{1:j'})\right)
\right)
}.
\label{eq:mult_factor_gibbs}
\end{split}
\end{equation}
So if we already have $\Pr(\boldsymbol{\pi}'' | \boldsymbol{\pi}_{(-j)})$, we can divide it by \eqref{eq:mult_factor_gibbs} to quickly obtain $\Pr(\boldsymbol{\pi}' | \boldsymbol{\pi}_{(-j)})$.
Moreover, we can efficiently calculate \eqref{eq:mult_factor_gibbs} by storing previous computational results: for instance, the summation over the risk group $R(\boldsymbol{\pi}_{1:j'}')$ shares many elements with the summation over the risk group $R(\boldsymbol{\pi}_{1:j'}'')$.
Similar ideas can be used to speed up other calculations required for MCEM.

\begin{algorithm}
	\caption{Gibbs sampler for mutation orders}
	\label{algo:gibbs}
	\begin{algorithmic}
		\STATE Initialize Gibbs step index $k = 1$ and mutation order $\boldsymbol{\pi}^{(0)}$.
		\FOR{Gibbs sweep index $i=1,2,...$}
			\FOR{$j \in \{1,...,n\}$}
				\STATE $\boldsymbol{\pi}_{(-j)} \coloneqq \boldsymbol{\pi}^{(k - 1)}_{(-j)}$
				\STATE Sample $\boldsymbol{\pi}^{(k)}$ from the distribution $\Pr \left (\boldsymbol{\pi} \big| \boldsymbol{\pi}^{(k - 1)}_{(-j)}\right )$.
				\STATE $k \coloneqq k + 1$
			\ENDFOR
		\ENDFOR
	\end{algorithmic}
\end{algorithm}

Given the Monte Carlo samples from the E-step, the M-step maximizes the mean log-likelihood of the complete data.
Suppose the E-step generates Monte Carlo samples $\boldsymbol{\pi}^{(1)},...,\boldsymbol{\pi}^{(E)}$.
Then during the M-step, we solve
\begin{align}
\label{eq:likelihood_observed_data_estep}
\max_{\boldsymbol{\theta}} \
\frac{1}{E}\sum_{i=1}^E
\log
\mathcal{L}_c
\left (
\mathbf{S}_{\text{obs}},
\boldsymbol{\pi}^{(i)}
;
\boldsymbol{\theta}
\right )
\end{align}
using iterative procedures such as gradient ascent.

We use ascent-based MCEM \citep{caffo2005ascent} to maintain the monotonicity property of the EM algorithm.
Briefly, ascent-based MCEM gives a rule for deciding if the proposed MCEM estimate at each iteration should be accepted or if the Monte Carlo sample size should be increased.
As the number of Monte Carlo samples increases, the standard error of the estimated expected log likelihood decreases.
So for a sufficiently large number of Monte Carlo samples, we can ensure that the observed data likelihood increases with high probability.

\subsection{Regularization and variable selection}
\label{sec:lasso}

In many cases, it is desirable to model the effects of many features.
For instance, \citet{Yaari2013-dg} estimate a 5-mer model with $1024$ parameters.
Estimating the parameters for a per-target model increases the number of parameters by an additional factor of four.
If the number of sequences in the dataset is small compared to the number of features, the optimization problem in \eqref{eq:likelihood_observed_data} can be ill-posed.
For such high-dimensional settings, it is common to use regularization to stabilize our estimates and encourage model structure.

In particular, we may believe that only a small subset of the features affects the mutation rate.
\citet{Yaari2013-dg} assume that the nucleotides closest to a position have the most significant effect on its mutation rate: for 5-mer motifs with a small number of observations, they estimate its mutation rate using an offset 3-mer motif.
In our method, we use the lasso \citep{tibshirani1996regression} to perform variable selection.

To incorporate the lasso, our estimation procedure requires two steps.
The first step maximizes the  observed log-likelihood with a lasso penalty and thereby performs variable selection.
The second step aims to quantify the uncertainty of our model parameter estimates: we refit the model parameters by maximizing the unpenalized objective and use the confidence intervals for the unpenalized model as an assessment of uncertainty.

In the first step, we split the data into training and validation sets denoted $\mathbf{S}_{\text{obs,train}}$ and $\mathbf{S}_{\text{obs,val}}$, respectively, and maximize the penalized log-likelihood of the training data
\begin{align}
\hat{\boldsymbol{\theta}} = \arg\max_{\boldsymbol{\theta}} \left\{ \log \mathcal{L}(\mathbf{S}_{\text{obs,train}} ; \boldsymbol{\theta})
- \lambda \|\boldsymbol{\theta}\|_1 \right\},
\label{eq:lasso_log_lik}
\end{align}
where $\lambda > 0$ is a penalty parameter. To solve \eqref{eq:lasso_log_lik}, we use a variant of MCEM: the E-step is the same as before, but we maximize the penalized EM surrogate function during the M-step. The penalized EM surrogate function is simply \eqref{eq:likelihood_observed_data_estep} with a lasso penalty:
\begin{align}
\frac{1}{E} \sum_{i=1}^E
\log \mathcal{L}_c
\left (
\mathbf{S}_{\text{obs,train}},
\boldsymbol{\pi}^{(i)}
; \boldsymbol{\theta}
\right )
- \lambda \|\boldsymbol{\theta}\|_1.
\label{eq:penalized_em_surr}
\end{align}
This can be maximized using the generalized gradient ascent algorithm given in Algorithm~\ref{algo:mstep_lasso} \citep{beck2009fast,nesterov2013gradient}.

\begin{algorithm}
	\caption{M-step via generalized gradient ascent}
	\label{algo:mstep_lasso}
	\begin{algorithmic}
		\STATE Initialize $\boldsymbol{\theta}$. Choose a step size $\alpha >0$.
		\FOR{iteration $k=1,2,...$ until convergence}
			\STATE
			$$
			\boldsymbol{\theta} \coloneqq \boldsymbol{\theta} + \alpha \nabla_{\theta}
			\frac{1}{E} \sum_{i=1}^E
			\mathcal{L}_c(\mathbf{S}_{\text{obs,train}}, \boldsymbol \pi^{(i)}; \boldsymbol{\theta} )
			$$
			\FOR{parameter index $j = 1,..,p$}
				\STATE $\theta_j \coloneqq \text{sign}(\theta_j)\max(|\theta_j - \lambda |, 0) $
			\ENDFOR
		\ENDFOR
	\end{algorithmic}
\end{algorithm}

We tune the penalty parameter $\lambda$ in \eqref{eq:lasso_log_lik} by training-validation split.
In the typical ideal case, we choose the penalty parameter that maximizes the likelihood of the observed validation data.
Unfortunately the likelihood of observed data is computationally intractable.
Instead we use the property that, for any $\boldsymbol{\theta}$ and $\boldsymbol{\theta}'$, the difference between the log-likelihoods of the observed data is bounded below by the difference between the expected log-likelihoods of the complete data
\begin{equation}
\begin{split}
\log \mathcal{L}(\mathbf{S}_{\text{obs}} ; {\boldsymbol{\theta}}) -
&\log \mathcal{L}(\mathbf{S}_{\text{obs}} ; {\boldsymbol{\theta}}' )
\ge\\
& \qquad \mathbb{E} \left[
\log \mathcal{L}_c(\mathbf{S}_{\text{obs}}, \boldsymbol{\pi}; {\boldsymbol{\theta}})
 - \log \mathcal{L}_c(\mathbf{S}_{\text{obs}}, \boldsymbol{\pi}; { \boldsymbol{\theta}}' )
 \mid
 \mathbf{S}_{\text{obs}}; \boldsymbol{\theta}'
 \right],
\label{eq:surrogate}
\end{split}
\end{equation}
which follows directly from Jensen's inequality.
The expectation above is taken with respect to the conditional distribution of the mutation orders $\boldsymbol{\pi}$ given the observed data $\mathbf{S}_{\text{obs}}$ and model parameter $\boldsymbol{\theta}'$.
Thus the right-hand side can be estimated by sampling mutation orders from the Gibbs sampler in Algorithm~\ref{algo:gibbs}.
If the right-hand side of \eqref{eq:surrogate} is positive, then $\boldsymbol{\theta}$ has a higher log-likelihood than $\boldsymbol{\theta}'$ on the validation set.
However, if the right-hand side is negative, we do not know how the two parameters compare.

Our proposal for tuning the penalty parameter, Algorithm~\ref{algo:tuning_lambda}, is based on \eqref{eq:surrogate}.
The algorithm searches across a one-dimensional grid of penalty parameters, from largest to smallest.
For consecutive penalty parameters, we estimate the right-hand side of \eqref{eq:surrogate} to determine if the smaller penalty parameter has a higher observed log-likelihood.
We keep shrinking the penalty parameter until the estimate for the right-hand side of \eqref{eq:surrogate} is negative.
Since the check based on \eqref{eq:surrogate} is conservative, we may end up choosing a penalty parameter that is slightly larger than desired.
Nonetheless, our simulation results suggest that this procedure works well in practice.

Algorithm~\ref{algo:tuning_lambda} can be easily extended to incorporate multiple training-validation splits such as in $k$-fold cross-validation: we average the estimates of the right-hand side of \eqref{eq:surrogate} across the training-validation splits and stop shrinking the penalty parameter if the average is negative.
After selecting a penalty parameter, we obtain the final parameter support from the $k$-fold procedure by refitting the penalized model on the whole training set.

\begin{algorithm}
	\caption{Tuning penalty parameters via training-validation split}
	\label{algo:tuning_lambda}
	\begin{algorithmic}
		\STATE Consider a grid of penalty parameters $\lambda_1 > ... > \lambda_K \ge 0$.
		\STATE Initialize $\lambda_{\text{best}} \coloneqq \lambda_1$. Fit $\lambda_1$ to get $\hat{\boldsymbol{\theta}}_{(1)}$.
		\FOR{iteration $i = 2,...,K$}
			\STATE Solve \eqref{eq:lasso_log_lik} with $\lambda \equiv \lambda_i$ to get $\hat{\boldsymbol{\theta}}_{(i)}$.
			\STATE Estimate via Monte Carlo
			\begin{align}
			\eta = \mathbb{E} \left[
			\log \mathcal{L}_c \left (
			\mathbf{S}_{\text{obs,val}}, \boldsymbol{\pi}
			;
			\hat {\boldsymbol{\theta}}_{(i )}
			\right )
			- \log \mathcal{L}_c \left (
			\mathbf{S}_{\text{obs,val}}, \boldsymbol{\pi}
			;
			\hat{ \boldsymbol{\theta}}_{(i -1)}
			\right )
			\big | \,
			\mathbf{S}_{\text{obs,val}}
			;
			\hat{\boldsymbol{\theta}}_{(i - 1)}
			\right].
			\label{eq:eta}
			\end{align}
			\IF{$\eta > 0$}
				\STATE $\lambda_{\text{best}} \coloneqq \lambda_i$
			\ELSE
				\STATE \textbf{break}
			\ENDIF
		\ENDFOR
	\end{algorithmic}
\end{algorithm}

Now we move on to the second step where our goal is to quantify the uncertainty of our estimated model parameters.
Unfortunately, estimating confidence intervals after model selection is a difficult problem, even in the much simpler case of linear models \citep{dezeure2015high}.
Hence some papers use the approach of fitting a penalized model, refitting an unpenalized model based on the selected variables, and then using the confidence intervals generated using traditional inference procedures for unpenalized models \citep{leeb2015various, hesterberg2008least}.
We proceed in the same manner: we refit the model by maximizing the unpenalized observed log-likelihood \eqref{eq:likelihood_observed_data} of the entire dataset with respect to the selected variables and constraining the others to zero; then we construct confidence intervals for the unpenalized model, ignoring the fact that we have already peeked at the data in the first step.
Though these confidence intervals are asymptotically valid only under very restrictive conditions, they provide some measure of the uncertainty of our fitted parameters; we show via simulation in Section~\ref{sec:simulations} that the coverage of these intervals is close to nominal.
To highlight that these intervals are not truly confidence intervals, we refer to them as uncertainty intervals, where 100(1 - $\alpha$)\% uncertainty intervals are constructed using intervals with nominal 100(1 - $\alpha$)\% coverage.

To obtain these uncertainty intervals, we calculate the standard error of our estimates using an estimate of the observed information matrix.
\citet{louis1982finding} shows that the observed information matrix is related to the complete data likelihood via the following identity:
\begin{equation}
\begin{split}
I\left[
\boldsymbol{\theta} \mid \mathbf{S}_{\text{obs}}
\right]
&
=
- \mathbb{E} \left[
\nabla_{\boldsymbol{\theta}}^2 \log \mathcal{L}_c(
\mathbf{S}_{\text{obs}}, \boldsymbol{\pi}
;
\boldsymbol{\theta}
)
\mid
\mathbf{S}_{\text{obs}};
\boldsymbol{\theta}
\right ] \nonumber \\
& \! -
\mathbb{E} \left[
\nabla_{\boldsymbol{\theta}}
\log \mathcal{L}_c(\mathbf{S}_{\text{obs}}, \boldsymbol{\pi}
;
\boldsymbol{\theta}
)
\left(
\nabla_{\boldsymbol{\theta}} \log \mathcal{L}_c(
\mathbf{S}_{\text{obs}}, \boldsymbol{\pi}
;
\boldsymbol{\theta} ) \right )^\top
\mid
\mathbf{S}_{\text{obs}};
\boldsymbol{\theta}
\right] \label{eq:louis} \\
& \! +
\mathbb{E} \left[
\nabla_{\boldsymbol{\theta}} \log \mathcal{L}_c(
\mathbf{S}_{\text{obs}}, \boldsymbol{\pi}
;
\boldsymbol{\theta})
\mid
\mathbf{S}_{\text{obs}};
\boldsymbol{\theta}
\right]
\mathbb{E}^\top\left[
\nabla_{\boldsymbol{\theta}} \log \mathcal{L}_c(
\mathbf{S}_{\text{obs}}, \boldsymbol{\pi}
;
\boldsymbol{\theta})
\mid
\mathbf{S}_{\text{obs}};
\boldsymbol{\theta}
\right]. \nonumber
\end{split}
\end{equation}
Therefore we can estimate the observed information matrix using samples from the final MCEM iteration and then invert it to obtain uncertainty intervals.

Finally, one caveat of our method is that the two-step procedure is not guaranteed to give estimates of standard errors/uncertainty intervals: The first step of our procedure may choose a penalty parameter such that the estimated information matrix in the second step is not positive definite.
We see this behavior in a small number of simulations in Section~\ref{sec:simulations}, though we do not observe such behavior in our data analysis.
To avoid this issue, we suggest combining $k$-fold cross-validation with Algorithm~\ref{algo:tuning_lambda} and use the average estimate of the lower bound \eqref{eq:eta} from each of the $k$ folds to tune the penalty parameter.

Our GPLv3-licensed Python implementation of \texttt{samm} is available at \url{http://github.com/matsengrp/samm}.
The repository includes code used for generating plots in this manuscript, as well as a tutorial for how to run \texttt{samm}.
All output from Sections~\ref{sec:simulations} and~\ref{sec:data} as well as the Appendix is available on \url{http://zenodo.org/record/1321330} with DOI \texttt{10.5281/zenodo.1321330}.

\subsection{Examples}
\label{sec:sim_examples}
By varying the motif dictionary $\M$, our procedure can fit different models of the mutation process.
In this section, we list some example models that can be fit using our procedure and discuss the interplay between the motifs included in $\M$ and our feature-selection step.
In the simplest case, analogous to existing work \citep{Yaari2013-dg,Cui2016-wz}, we can estimate a ``$k$-mer model'' (where $k$ is odd) by letting
\begin{align}
\M = \left\{(m, (k +1)/2) : m \in \{\A, \C, \G, \T\}^k \right\}.
\label{eq:kmer}
\end{align}
The lasso would encourage setting elements in $\boldsymbol{\theta}$ to zero, which means that these $k$-mer motifs would have the same baseline risk of experiencing a mutation.

In practice, instead of modeling only the effects of $k$-mers for a fixed $k$, we may believe that the hazard rate for a position is affected more by positions closer to it.
In this case, we can model the effect of $z$-mers of varying length, e.g., $1,3,...,k$-mer motifs, with
\begin{align}
\M = \left\{(m, (z +1)/2) : m \in \{\A, \C, \G, \T\}^z, z \in 1,3,\ldots k \right\}.
\label{eq:hier_kmer}
\end{align}
We refer to this model as ``hierarchical'', as the elements in $\M$ relate to each other in a nested fashion.
By including motifs in a hierarchical fashion, the lasso penalty encourages $z$-mers with the same inner $(z - 2)$-mer to share the same mutation rate.
This model formalizes the intuition used by \citet{Yaari2013-dg}: they try to estimate the mutation rates of 5-mers but fall back to using a 3-mer sub-motif if that 5-mer does not appear enough times in the data.

As mentioned before, we can add offset motifs to our motif dictionary as previous work suggests the mutation rates depend on upstream or downstream motifs \citep{Rogozin1992-xv,Pham2003-jm,Yaari2013-dg}.
For instance, one can include all the offset motifs that overlap the mutating position in the motif dictionary.
We refer to such models as offset $k$-mer models.

Finally, we can model the hazard rate of motifs mutating to different target nucleotides as in \eqref{eq:hazard_rate_target}.
We parameterize the model using $\boldsymbol{\theta}$ and $\boldsymbol{\theta}_\N$ for $\N \in \{ \A, \C, \G, \T \}$ since the penalized per-target model
\begin{equation}
\begin{split}
\arg\max_{\boldsymbol{\theta}, \boldsymbol{\theta}_{\N}: \N\in\{\A,\C,\G,\T\}}
\log \mathcal{L}\left(
\mathbf{S}_{\text{obs,train}} ; \right. & \left. \boldsymbol{\theta}, \{\boldsymbol{\theta}_{\N}: \N\in\{\A,\C,\G,\T\}\}
\right )\\
& - \lambda \left(\|\boldsymbol{\theta}\|_1 + \sum_{\N\in\{\A,\C,\G,\T\}} \|\boldsymbol{\theta}_{\N}\|_1 \right )
\end{split}
\end{equation}
will encourage hazard rates for the different target nucleotides to be the same if they share the same motif.

Many of these example models are overparameterized in order to obtain some desired sparsity pattern.
Such overparameterized models may have singular information matrices during the refitting procedure.
However this is not a problem since we are truly interested in the confidence intervals for the parameters $\boldsymbol{\theta}_{\text{agg}} = \boldsymbol{A} \boldsymbol{\theta}$ associated with the simple $k$-mer model, where $\boldsymbol A$ is a matrix that aggregates hierarchical motifs into a single $k$-mer.
Since this aggregate $k$-mer model is identifiable, we can get uncertainty intervals for $\boldsymbol{\theta}_{\text{agg}}$:
we calculate the pseudo-inverse $\boldsymbol I^-$ of the (estimated) information matrix and then use $\boldsymbol A \boldsymbol I^-\boldsymbol A^\top$ to get an estimate of the covariance matrix of $\boldsymbol{\theta}_{\text{agg}}$.

\section{Simulation results}\label{sec:simulations}

We now present a simulation study of our procedure, including a comparison to the current state-of-the-art method \texttt{SHazaM} \citep[version 0.1.8]{Yaari2013-dg} and the logistic regression approach in Section~\ref{sec:logistic}.

\subsection{Understanding the effect of various models and settings}\label{sec:simulations1}
We fit the following three models to simulated data:
\begin{itemize}
	\item 3-mer model: the hazard rate modeled by \eqref{eq:hazard_rate} with motif dictionary \eqref{eq:kmer} where $k = 3$,
	\item 3-mer per-target model: the hazard rate modeled by \eqref{eq:hazard_rate_target} with motif dictionary \eqref{eq:kmer} where $k = 3$,
	\item 2,3-mer model: the hazard rate modeled by \eqref{eq:hazard_rate} with motif dictionary
		$$
		\M = \left\{(m, j') : m \in \{\A, \C, \G, \T\}^2, j' \in \{1,2\}\right\} \cup \left\{(m, 2) : m \in \{\A, \C, \G, \T\}^3\right\}.
		$$

\end{itemize}
To understand how dataset composition affects the performance of our procedure, we simulate different datasets by varying the sample sizes, sparsity levels, and effect sizes.

We generate the true $\boldsymbol{\theta}^*$ according to the same hierarchical structure as each model we consider.
Let the model parameters corresponding to the motif $m$ be $\theta_m^*$ and corresponding to motif $m$ with target nucleotide $\N$ be $\theta_{m \rightarrow \N}^*$.
To obtain the desired sparsity level, we randomly select a portion of the parameters to zero out.
For per-target parameters, instead of setting the probability of mutating to $\N$ to zero, we set $\theta_{m \rightarrow \N}^*$ to $\log 1/3$ for all possible values of $\N$, indicating no mutation preference.
We scale the model parameters appropriately to control the effect size.

Our goal with these simulations is to obtain synthetic data that reflects different possible settings one may encounter when analyzing experimental data.
We use the experimental data in \citet{Cui2016-wz} analyzed in Section~\ref{sec:data} as a template and alter various underlying properties of this dataset to simulate data that replicates what typical real-world datasets look like.
We first generate na\"ive sequences using \texttt{partis}\footnote{Version 0.12.0: \url{http://git.io/fNvOx}} \citep{ralph2016consistency,ralph2016likelihood} by drawing a set of genes from the IMGT database \citep{lefranc1999imgt} and simulating an observation frequency for each.
Antibodies are composed of two units, a heavy and a light chain.
Further, light chains can be classed as either $\kappa$ or $\lambda$ depending on where the encoded sequence came from in the genome.
Both mice and humans have antibodies structured in this way.
We select only $\kappa$-light chain mouse BCRs for our simulation, as this reflects our experimental data in Section~\ref{sec:data}.
To generate the true $\boldsymbol{\theta}^*$ parameters, we randomly draw values from the mouse somatic hypermutation targeting model \texttt{MK\_RS5NF} of \citet{Cui2016-wz}; we refer to these parameters as $\boldsymbol{\theta}_{\MK}^*$.
The \texttt{MK\_RS5NF} model is a collection of mutabilities and substitution probabilities from a 5-mer fit to $\kappa$-light chain mouse BCR data.

The average length of the na\"ive sequences is around 290 nucleotides.
We use the survival model to mutate between 1\% and 5\% of the positions of each na\"ive sequence, obtaining a collection of simulated BCR sequences.
Conditional on their na\"ive sequences, BCR sequences mutate independently.

We vary sparsity, effect size, and sample size as follows.
We generate the true $\boldsymbol{\theta}^*$ parameters with 25\%, 50\%, and 100\% non-zero elements.
We also consider different effect sizes by scaling $\boldsymbol{\theta}^*$ such that its variance is $50\%$, $100\%$, and $200\%$ of the variance of the values in $\boldsymbol{\theta}_{\MK}^*$.
Finally, we fit the model using 100, 200, and 400 mutated BCR sequences.
For the main manuscript, we report the simulation settings where we vary one simulation setting and fix the other settings to the middle value (e.g., we vary number of samples but keep the effect size at $100\%$ and the number of non-zero elements at $50\%$); we report the result from running one hundred replicates for each setting.
For the remaining possible settings, as each separate model fit takes on average an hour to complete, we run only ten replicates and report the results in the Appendix Section~\ref{sec:all_the_others}.

To determine the optimal penalty parameter for \texttt{samm}, we split the data by gene subgroups, an externally-defined categorization that groups genes that share at least 75\% identity at the nucleotide level \citep{Lefranc2014-gz}, reserving 20\% of subgroups for validation and the remainder for training.
Splitting by gene subgroup ensures that the training and validation sets look sufficiently different; otherwise the sequences in the validation set look nearly identical to those in the training set, and we select a penalty parameter that is too small.
We then apply Algorithm~\ref{algo:tuning_lambda} over a decreasing sequence of penalty parameter-values $10^{-j}, 10^{-(j + 0.5)}, 10^{-(j + 1)}, \ldots$.
The starting value for the sequence of penalty parameter values was pre-tuned so that we use a smaller $j$ for smaller effect sizes and sample sizes.
In particular, we chose $j = 1$ if effect size is $50\%$ or sample size is $100$; $j = 2$ if the effect size is $200\%$ or sample size is $400$; and $j = 1.5$ otherwise.

For each penalty parameter-value, we run a maximum of ten MCEM iterations.
Mutation orders are sampled from each Gibbs sampler run every eight sweeps, after an initial burn-in period of 16 Gibbs sweeps.
For each E-step, we sample four mutation orders and continue to double the number of sampled mutation orders if the proposed estimate is not accepted by ascent-based MCEM.
Once we have an estimate of the support of our model, we refit an unpenalized model to obtain uncertainty estimates.
We run MCEM until the model has converged and the variance estimates of the estimated model parameters are all nonnegative.

We assess the performance of our procedure using three measures.
These performance metrics are all calculated with respect to the aggregate model since our complete model is overparameterized by design.
We calculate the relative $\boldsymbol{\theta}$ error, defined as $\|\boldsymbol{\theta}-\boldsymbol{\theta}^*\|_2/\|\boldsymbol{\theta}^*\|_2$, to see how close the estimated parameters are to the true parameter $\boldsymbol{\theta}^*$.
We also calculate Kendall's tau coefficient to see how well our procedure ranks the motifs in terms of their mutabilities.
Finally we calculate the coverage of our approximate 95\% uncertainty intervals.
We define the average coverage as the proportion of aggregate model parameters where the uncertainty intervals covered the true value.
The coverage calculations only involve aggregate parameters not zeroed out by our models.

\begin{figure}
    \centering
    \includegraphics[width=\textwidth]{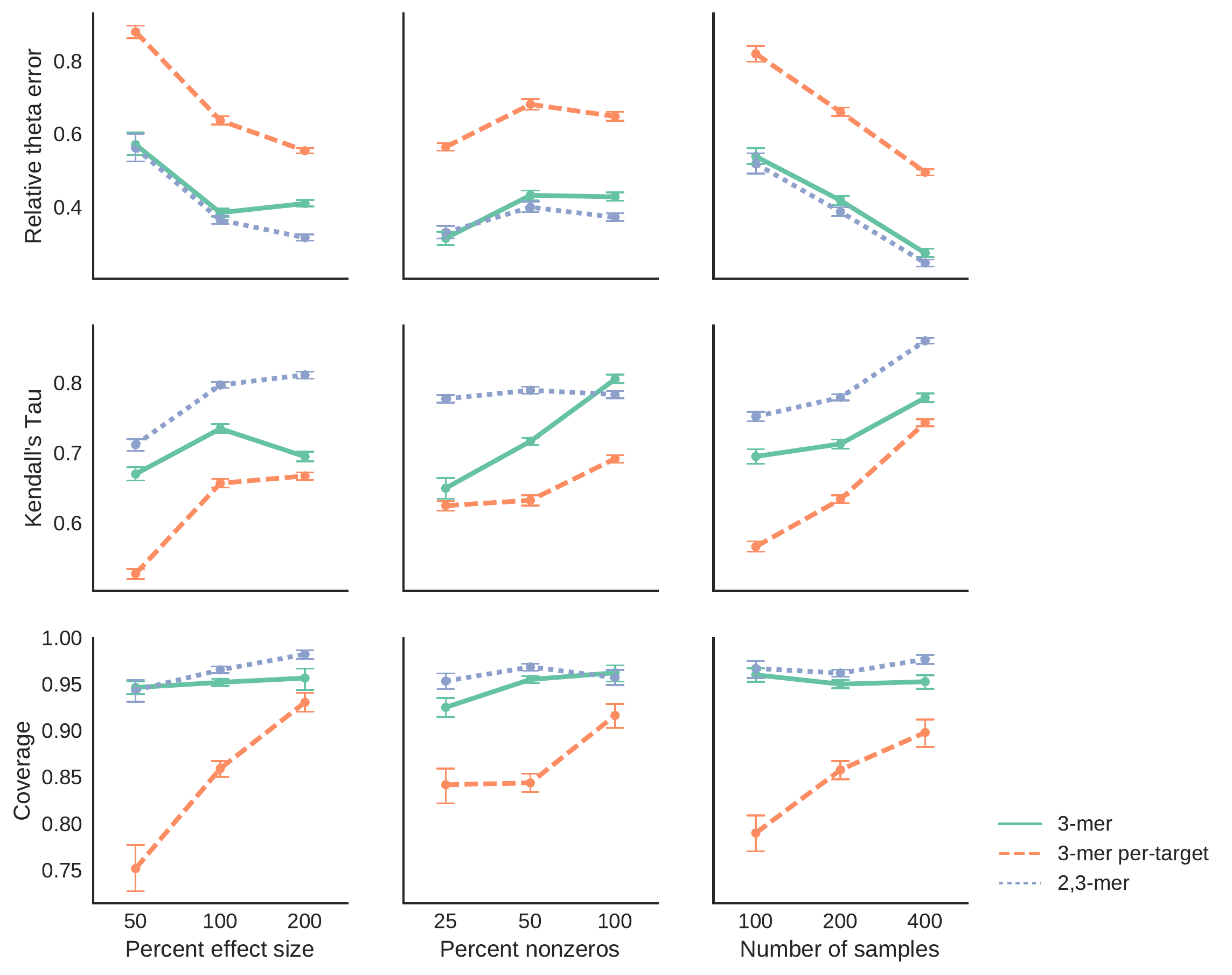}
    \caption{
        Relative error, correlation and coverage under different simulations settings for 3-mer, 3-mer per-target and 2,3-mer models.
    }
    \label{fig:simulation_basic}
\end{figure}

These simulations demonstrate that our estimation procedure performs as expected (Figure~\ref{fig:simulation_basic}).
As the sample size and effect size increase, the relative $\boldsymbol{\theta}$ error decreases and the rank correlation increases.
On the other hand, as the percent of non-zero elements increases, both the relative $\boldsymbol{\theta}$ error and rank correlation increase.
The error increases because there are more parameters to estimate.
The increase in rank correlation is likely an artifact of how the metric is calculated, as Kendall's tau removes ties from the calculations.
In particular, as the percent of non-zero elements increases, the number of ties in the data decreases, so the rank correlation seems to increase.
In all the plots, we see that the 3-mer per-target model tends to be the most difficult to estimate.
This is expected as it contains 256 parameters whereas the 3-mer model only has 64 parameters.

Our simulations show that the coverages for the 3-mer and the 2,3-mer models are close to 95\%, which is surprising as our uncertainty intervals ignore the double-peeking issue (Figure~\ref{fig:simulation_basic}).
\citet{zhao2017defense} explain why this procedure might work: under certain assumptions, the variables selected by the lasso are deterministic with high probability, so using the lasso to select variables does not really constitute as peeking at the data twice.

However, the coverage of the 3-mer per-target is much lower, dropping below 80\% in certain settings (Figure~\ref{fig:simulation_basic}).
We suspect that the low coverage is mainly due to a lack of data, as the coverage improves with the number of samples.
When there is a small number of samples compared to the number of parameters, our method may only provide a reasonable ranking of how mutable the motifs are but may not provide good estimates and uncertainty intervals.

Across the 2700 simulation runs, there were twenty where the estimated information matrices were not positive definite and therefore uncertainty intervals cannot be calculated (Table~\ref{tab:failed_runs}).
We believe that this occurs when the selected penalty parameter is too small; for small penalty parameters, the support of the fitted model becomes too large.
In this case, when we refit the model with no penalty parameter the problem is ill-posed and therefore the estimated information matrix is not positive definite.
To avoid this issue, we recommend using $k$-fold cross-validation in practice, rather than just a training/validation split.
(We use 5-fold cross-validation for the real data analysis and do not run into this issue.)

\subsection{Method comparisons}
In this section, we compare the performance of \texttt{samm} to \texttt{SHazaM} and penalized logistic regression on simulated data.
Since \texttt{SHazaM} only estimates the effect of 5-mer motifs, we simulate data such that the mutation rate at a specific site depends on the 5-mer centered at that position and the target nucleotide.
We simulate 2000 BCR sequences from 4 mice.
For each mouse, we generate a separate set of na\"ive sequences using the same procedure as in Section~\ref{sec:simulations1}.
From these na\"ive sequences, we simulate the mutation process independently to generate BCR sequences.
We use two methods to simulate the mutation process:
\begin{itemize}
	\item \textbf{Survival Simulation}: We generate model parameters $\boldsymbol{\theta}$ by resampling the values from $\boldsymbol{\theta}_{\MK}$ into a 3,5-mer per-target model structure.
    We then mutate the na\"ive sequences according to the survival model.
	\item \textbf{\texttt{SHMulate} Simulation}: We use $\boldsymbol{\theta}_{\MK}$ and mutate the na\"ive sequences using the \texttt{SHMulate} function in the \texttt{SHazaM} package \citep{Yaari2013-dg, gupta2015change}.
	\texttt{SHMulate} simulates the mutation process using a procedure that is similar to a survival model.
    However the exact calculations differ somewhat (e.g.\ it does not allow the mutation process to create stop codons).
\end{itemize}
\texttt{SHazaM} should have an advantage in the \texttt{SHMulate} simulations since the $\boldsymbol{\theta}_{\MK}$ was estimated using \texttt{SHazaM} on a separate BCR dataset and \texttt{SHazaM} uses some prior assumptions about the model structure.
In particular, \texttt{SHazaM} assumes that 5-mer motifs that share certain upstream/downstream nucleotides have similar mutabilities.
The simulations are run until 1--5\% of the sequence is mutated.
This mutation rate is on the low end for affinity-matured BCR sequences \citep[compare the $3\times$ higher rate in][]{He2014-fn}, giving \texttt{SHazaM} and logistic regression a slight edge since the mutation rates will not change for most positions with accumulation of BCR mutations.

We fit a 3,5-mer per-target \texttt{samm} model using the same procedure as in Section~\ref{sec:simulations1}.
Using the same motif dictionary, we also fit a 3,5-mer per-target logistic regression model using logistic regression with a lasso penalty.
We measure model performance by the relative $\boldsymbol{\theta}$ error and rank correlation over 100 simulation replicates.

Our method implemented in \texttt{samm} significantly outperforms logistic regression and \texttt{SHazaM} in both scenarios (Table~\ref{tab:5mer_sim}), even though \texttt{SHazaM} should have an advantage when we simulate data using a dense model from \texttt{SHMulate}.
Logistic regression and \texttt{SHazaM} tended to produce similar estimates, though logistic regression tended to do better when we simulated using the survival model and \texttt{SHazaM} tended to do better when we used the \texttt{SHMulate} model.

\begin{figure}
    \begin{tabular}{cc}
        \includegraphics[width=.49\textwidth]{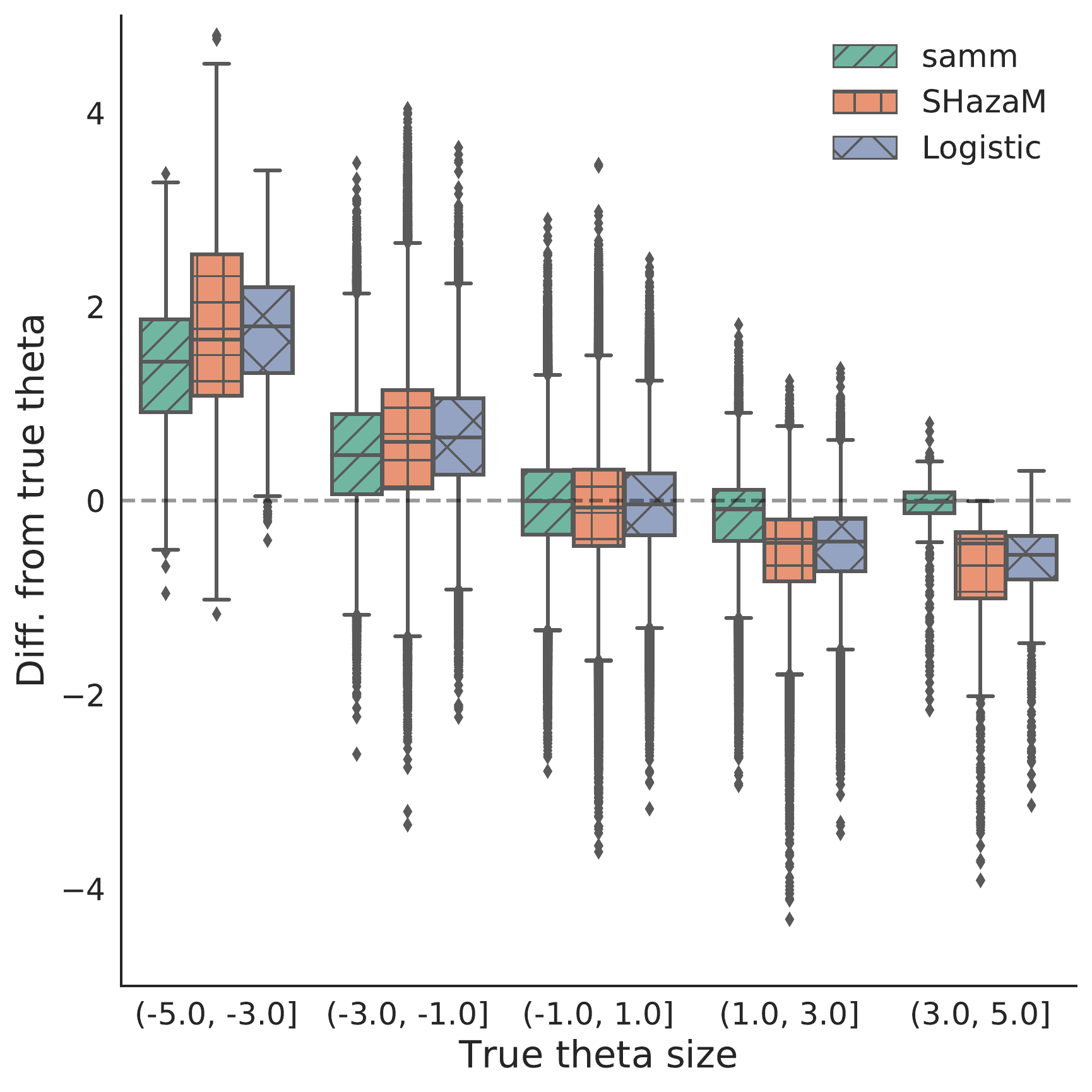}
        & \includegraphics[width=.49\textwidth]{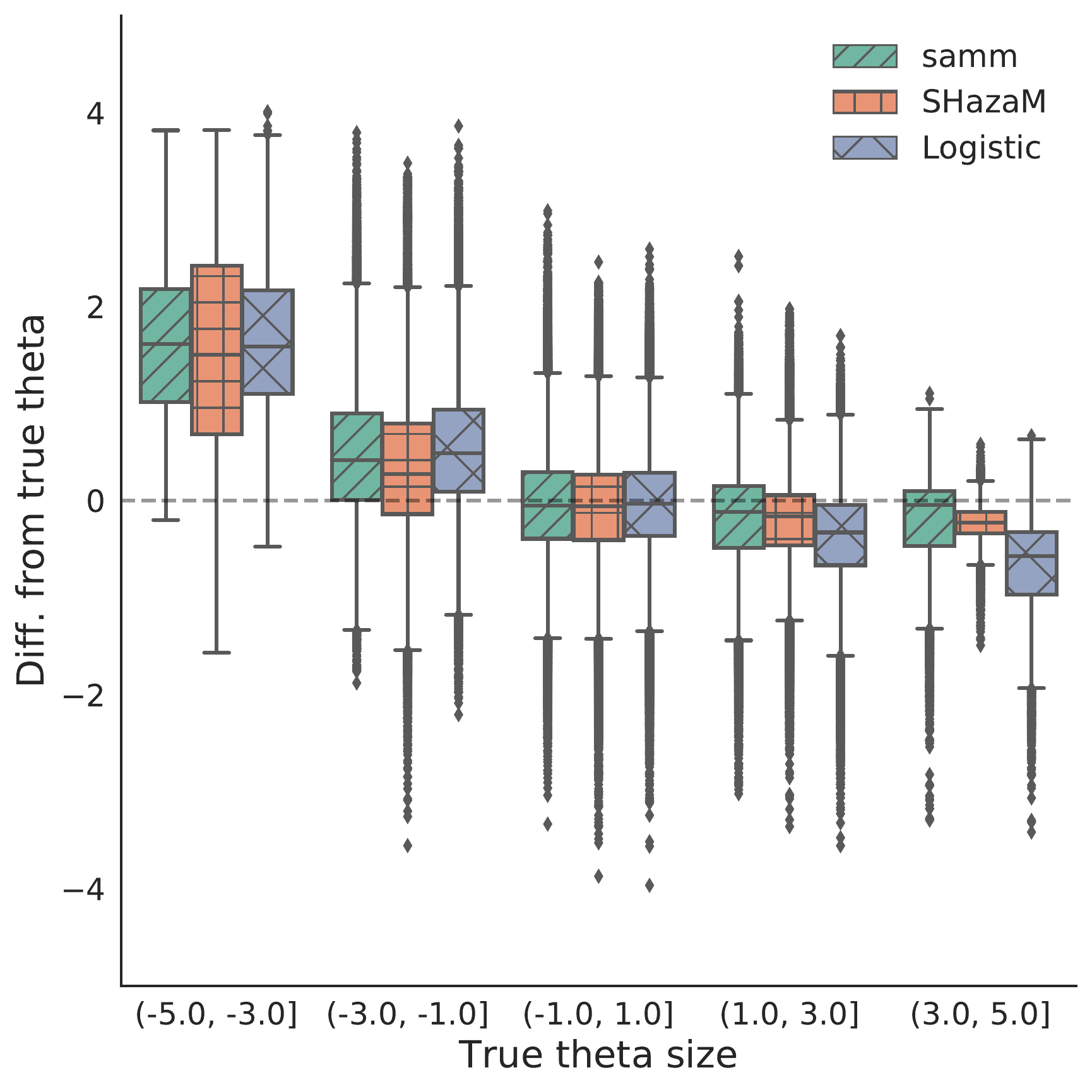}\\
        (a) Survival simulation & (b) \texttt{SHMulate} simulation
    \end{tabular}
    \caption{
        Boxplots of the differences between median-centered fitted and true $\boldsymbol{\theta}$ values for samm (left), \texttt{SHazaM} (middle), and logistic regression (right).
        }
    \label{fig:5mer_sim}
\end{figure}

We present the results of model fitting in more detail in Figure~\ref{fig:5mer_sim}.
For negative $\boldsymbol{\theta}$ values, all the methods are biased towards zero, though \texttt{SHazaM} and logistic regression tend to be more so.
For positive $\boldsymbol{\theta}$ values, \texttt{samm} is nearly unbiased while \texttt{SHazaM} and logistic regression are somewhat biased towards zero.
The methods probably have trouble estimating negative values since we only observe a small number of mutations per sequence and the data is more informative for finding motifs with high mutation rates rather than those with low mutation rates.
Based on results from Section~\ref{sec:simulations1}, we expect the bias of \texttt{samm} to shrink as the number of training observations increases.

\begin{table}
\caption{
	Comparison of \texttt{samm}, \texttt{SHazaM}, and penalized logistic regression given 2000 simulated B-cell receptor sequences from 4 mice.
    Relative $\boldsymbol\theta$ error and Kendall's tau computed separately for each of the 100 replicates.
	Monte Carlo standard errors calculated over these 100 estimates are given in parentheses.}
\label{tab:5mer_sim}
\begin{tabular}{rrrr}
    \toprule
    Simulator & Model & Relative $\boldsymbol{\theta}$ error & Kendall's tau \\
    \midrule
    \multirow{2}{*}{survival model} & \texttt{samm}   & 0.571 (0.002) & 0.630 (0.001) \\
    & \texttt{SHazaM} & 0.731 (0.002) & 0.507 (0.002) \\
    & logistic  & 0.611 (0.002) & 0.596 (0.001) \\
    \midrule
    \multirow{2}{*}{\texttt{SHMulate}} & \texttt{samm}   & 0.478 (0.001)  & 0.689 (0.001) \\
    & \texttt{SHazaM} & 0.489 (0.001) & 0.690 (0.001)\\
    & logistic  & 0.499 (0.002)  & 0.677 (0.001)
\end{tabular}
\end{table}

\section{Data analysis}\label{sec:data}

We fit models to the BCR sequence data obtained from a vaccination study of four transgenic mice published in \citep{Cui2016-wz}.
In this experimental setting, the substitutions present in the $\kappa$-light chain sequences are unlikely to be affected by natural selection on BCR function.
Thus we restrict our analysis to only $\kappa$-light chain data in order to estimate somatic hypermutation rates, rather than a combination of somatic hypermutation and selection \citep{Yaari2012-kk,McCoy2015-qi,Yaari2015-ss}.
A single na\"ive sequence can give rise to many different B-cell receptors by somatic hypermutation, forming a so-called ``clonal family'' which may have varying levels of shared evolutionary history.
We use \texttt{partis} \citep{ralph2016consistency} to assign mutated sequences to clonal families and infer the most likely na\"ive sequence in each family.
In both the sequencing and the clonal family inference there is the possibility of error propagation; we begin our analysis by assuming BCRs are accurately sequenced and assigned to clonal families.
The resulting data has the composition shown in Table~\ref{tab:data-stats}.
To mitigate double-counting mutations, we sample a single mutated sequence from each clonal family.
Though this discards a lot of the data, we believe this gives more accurate estimates than other approaches that try to use all the data or estimate mutation history; we analyze this issue in more depth in Section~\ref{appendix:parsimony} in the Appendix.

\begin{table}[h!]
    \caption{
        Statistics of processed $\kappa$-light chain data from \citet{Cui2016-wz}.
        \texttt{SHazaM} uses all sequences while \texttt{samm} samples a single sequence from each clonal family.
        We filter sequences with indels in all analyses.
        There are fewer clonal families in the sampled sequences as \texttt{samm} filters out sequences with no mutations.
    }
    \label{tab:data-stats}
    \begin{tabular}{lrr}
        \toprule
        & All sequences & Sampled sequences\\
        \midrule
        Number of mutated sequences                 &  15,025 & 2,429\\
        Number of clonal families                   &  2,565  & 2,429\\
        Median mutated sequence length              &  282    & 282\\
        Average mutation frequency (\%)             &  2.32   & 2.17\\
        Number of 5-mers in na\"ive sequences         &  1,014  & 967\\
        \bottomrule
    \end{tabular}
\end{table}

\begin{figure}
	\centering
	\includegraphics[width=\textwidth]{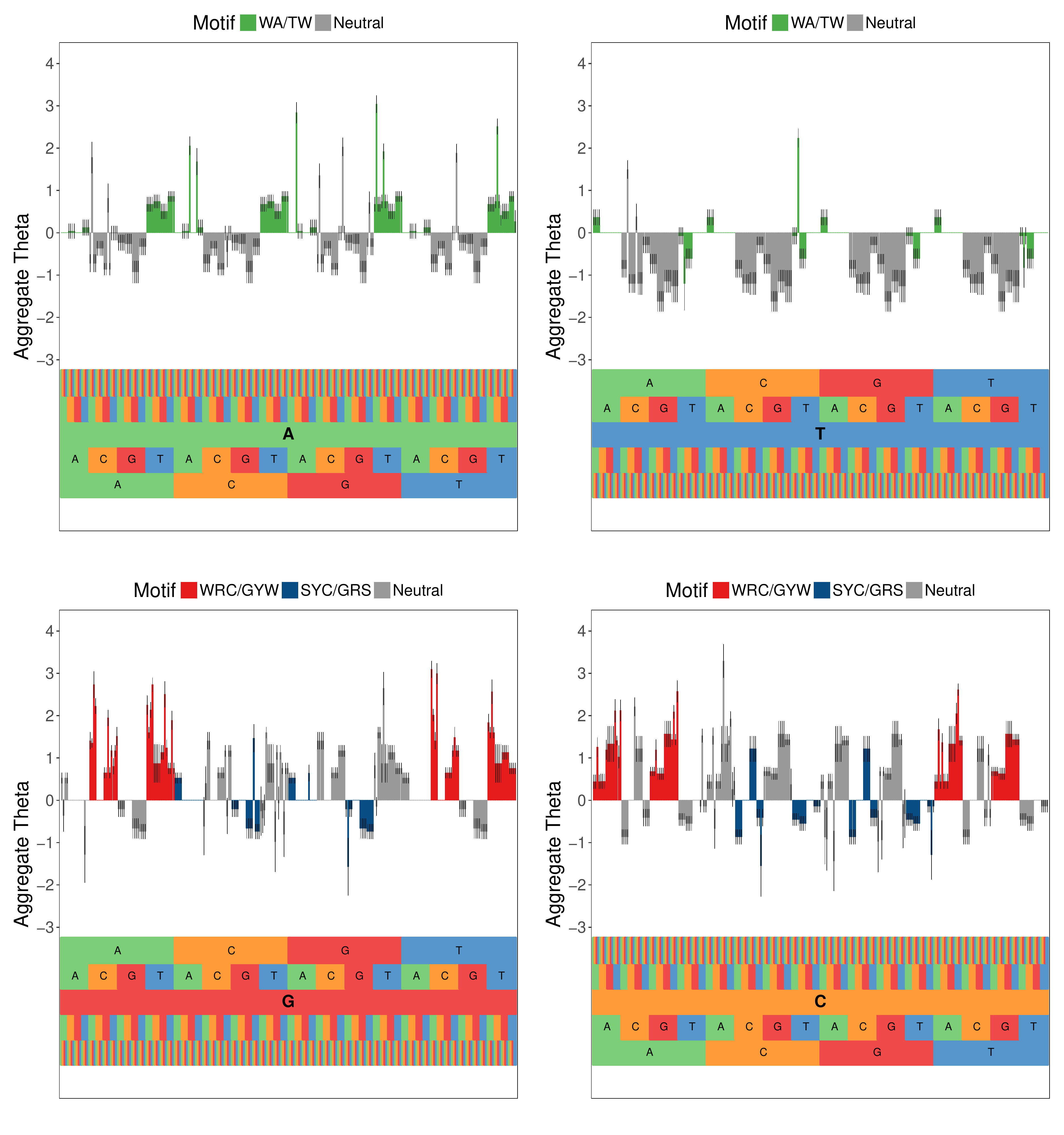}
	\caption{
		Estimated somatic hypermutation model for mouse light chains using \texttt{samm} for 5-mer motifs centered on the bases \A\ (top left), \T\ (top right), \G\ (bottom left), and \C\ (bottom right).
		The motif corresponding to an $x$-axis position can be read from bottom to top.
		Plots depict the estimated aggregate $\theta$ of 5-mer motifs after estimating the model for a 3,5-mer model and aggregating estimates using the procedure outlined in Section~\ref{sec:sim_examples}.
        A negative value means a reduced mutation rate relative to the baseline hazard, whereas a positive means an enhancement.
        Well-known hot spots, $\W\R\underline{\C}/\underline{\G}\Y\W$ and $\W\underline{\A}/\underline{\T}\W$, are colored red and green, respectively.
		The well-known cold spot $\Snuc\Y\underline{\C}/\underline{\G}\R\Snuc$ is colored blue.
		All other motifs are colored grey.
		The 95\% uncertainty intervals for the estimates are depicted by black lines in the center of each bar.
	}
	\label{fig:hedgehog}
\end{figure}

We fit a 3,5-mer model using \texttt{samm} using the same settings as before (Figure~\ref{fig:hedgehog}), though with 5-fold cross-validation to determine the optimal parameter support.
The $\boldsymbol{\theta}$ estimate has a block-like and 4-fold-repetitive pattern because many 5-mer motifs were zeroed out during the lasso step.
The 95\% uncertainty intervals suggest that many motifs have a marked nonzero effect.

Our model recovers many of the well-known ``hot'' (more mutable) and ``cold'' spots (less mutable $k$-mers), which are denoted by the red, blue, and green bars in Figure~\ref{fig:hedgehog}.
Hot/cold motifs are typically denoted with an underline indicating which position is mutating and represented by degenerate bases $\W=\{\A, \T\}$, $\R=\{\A, \G\}$, $\Y=\{\C, \T\}$, $\Snuc=\{\C, \G\}$, $\N=\{\A, \G, \C, \T\}$.
We confirm that many highly mutable 5-mer motifs match the classical hot spot motif $\W\R\underline{\C}$ and its reverse complement $\underline{\G}\Y\W$ (since the mutation process can happen on either DNA strand) \citep{Rogozin2004-np}.
We also confirm that many less mutable 5-mer motifs match the canonical cold spot $\Snuc\Y\underline{\C}/\underline{\G}\R\Snuc$ \citep{Yaari2013-dg}.
For example, one of the 5-mers we estimate to have high mutability ($\theta=1.688$) is $\A\A\underline{\G}\C\T$, which is of the form $\N\N\underline{\G}\Y\W$ and ends with the 3-mer $\underline{\G}\Y\W$.
As $\C$ is an example of a $\Y$ nucleotide and $\T$ is an example of a $\W$, $\A\A\underline{\G}\C\T$ is an example of the hot spot motif $\underline{\G}\Y\W$.

Our model also reveals shortcomings with the current hot and cold spot definitions.
Our estimates show significant variability in the mutabilities of motifs, even if they contain the same hot or cold spot motif.
For instance, in the established literature the $\A\T\underline{\G}\G\C$ motif is considered to be a cold spot since it is of the form $\underline{\G}\R\Snuc$.
We estimate its $\theta$ value to be very large ($\theta=2.206$) relative to the other $\theta$ values, suggesting that it is actually a hot spot.
We also see \texttt{SHazaM} estimates all motifs of the form $\C\C\underline{\C}\N\N$ to have negative mutability, and these are examples of the known cold spot $\Snuc\Y\underline{\C}$.
Estimates from \texttt{samm} show $\C\C\underline{\C}\G\N$ has a positive mutability even though it is also of the form $\Snuc\Y\underline{\C}$, indicating the inner 3-mer $\C\underline{\C}\G$ may increase mutation rate more than the two $\C$ nucleotides to the left of the mutating position.
In addition, the classic hot spots with a central \T\ nucleotide actually had very low mutability estimates; this suggests that using the well-known $\W\underline{\A}/\underline{\T}\W$ to identify hot spots may not be appropriate.

Finally, our model suggests that \texttt{samm} can be used to discover new hot and cold spots.
For example, consider motifs with the central base $\C$ mutating.
We find that the mutabilities of the 5-mer $\C\A\underline{\C}\G\C$ and of the 3-mers $\G\underline{\C}\G$, $\G\underline{\C}\T$, $\A\underline{\C}\T$, and $\A\underline{\C}\G$ are all higher than any motif of the form $\W\R\underline{\C}$.
As each of these motifs are of the form $\N\R\underline{\C}$, this indicates the $\R$ nucleotide immediately preceding the mutating $\C$ may affect mutation rate more than the $\W$ nucleotide two bases away.
A well-defined inferential procedure to determine significant collections of hot and cold spots with ample support from the data will require additional future work.

\begin{figure}
	\centering
    \begin{tabular}{ccc}
	\centering
		\includegraphics[width=0.33\textwidth]{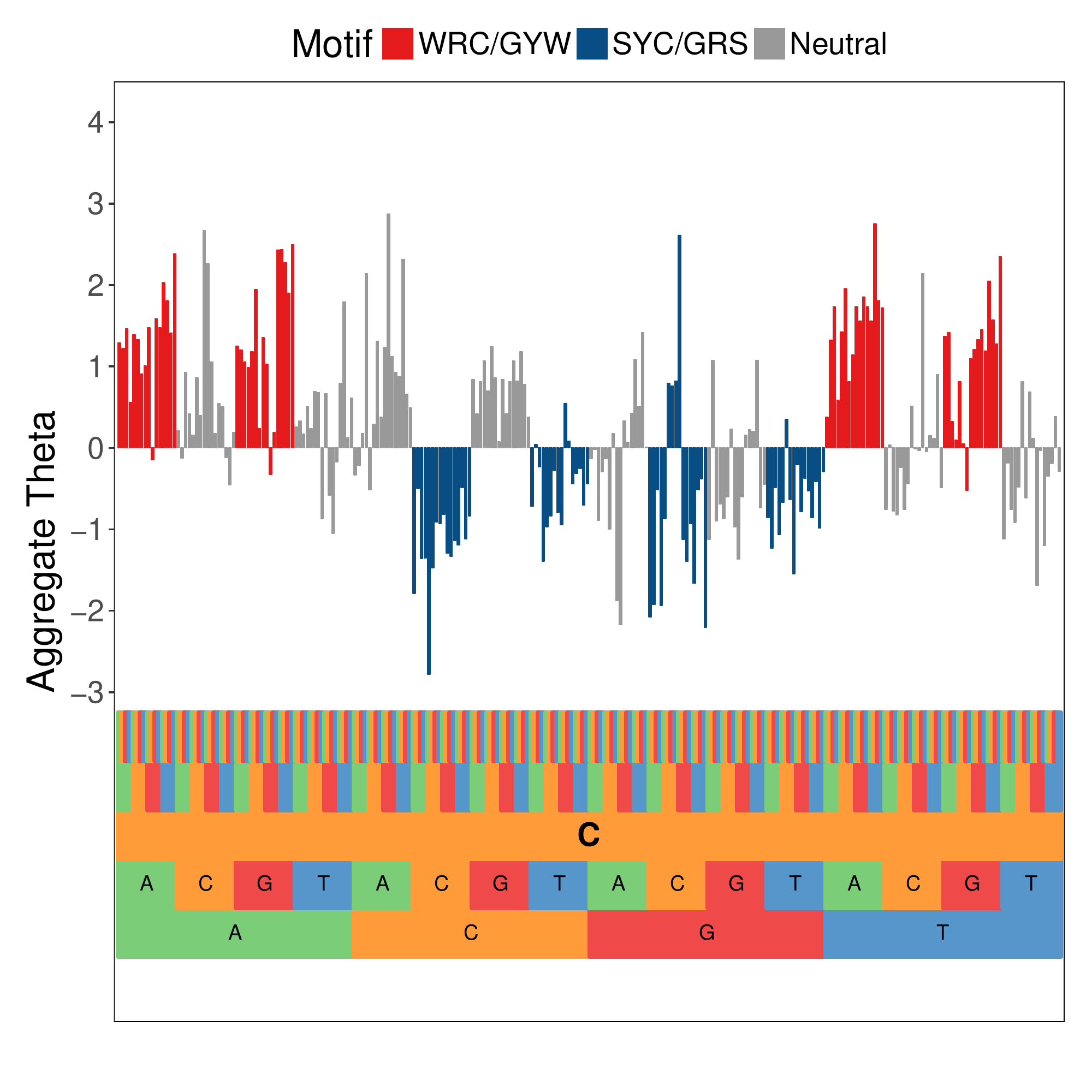} &
		\includegraphics[width=0.33\textwidth]{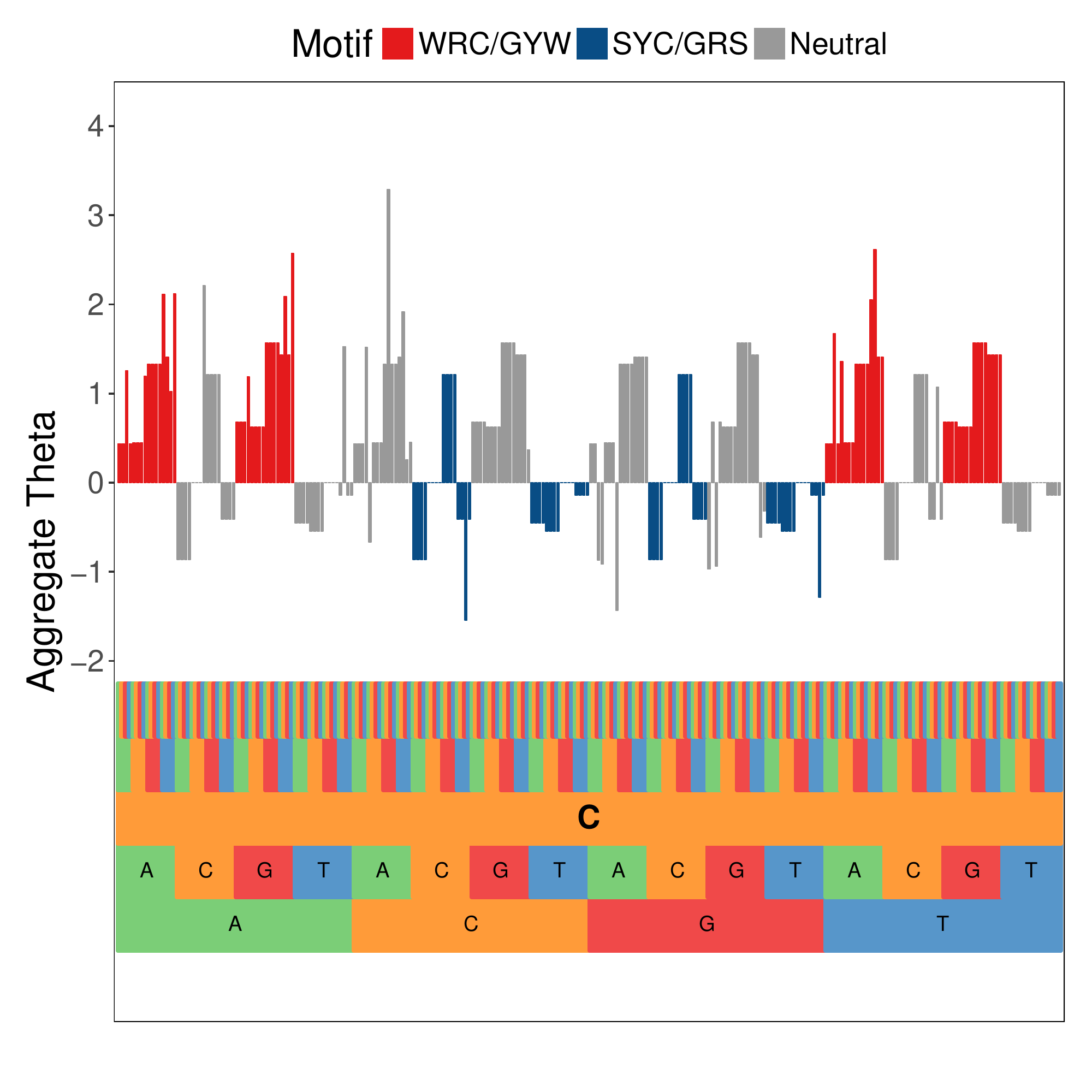} &
		\includegraphics[width=0.33\textwidth]{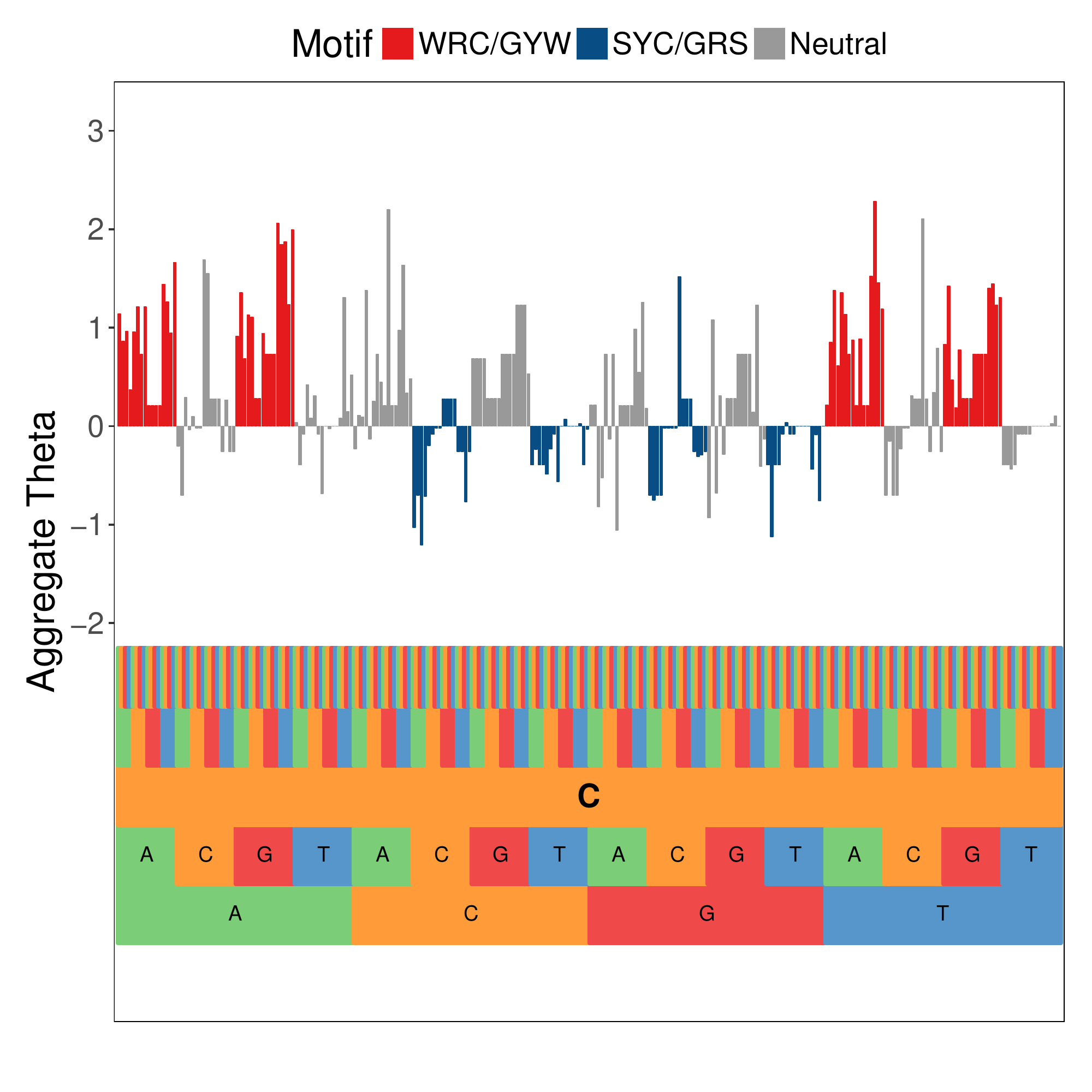}\\
		(a) \texttt{SHazaM}: 1015 unique $\theta$ values &
        (b) \texttt{samm}: 137 unique $\theta$ values &
		(c) logistic: 485 unique $\theta$ values
	\end{tabular}
	\caption{
		A comparison of fitted aggregate $\theta$ values from \texttt{SHazaM} (left), \texttt{samm} (middle), and logistic regression (right) for 5-mer motifs with central base \C.
		The \texttt{samm} fit is the same as in Figure~\ref{fig:hedgehog}.
        Both \texttt{samm} and logistic are 3,5-mer fits aggregated into 5-mer models.
		\texttt{samm} and logistic tend to fit more parsimonious models compared to \texttt{SHazaM}, so the left plot looks more ``spiky'' than the middle and right ones.
		\texttt{samm} produces the most parsimonious fits among the three methods.
	}
	\label{fig:shazam_v_samm_v_logistic}
\end{figure}

For comparison, we fit \texttt{SHazaM} on the same data without sampling a single sequence from each clonal family, as was done by \citet{Yaari2013-dg}.
We also fit the logistic model on the same data as \texttt{samm}.
All models use the data to determine the degrees of freedom to use in fitting $\boldsymbol{\theta}$, resulting in the number of unique $\boldsymbol{\theta}$ values fit to be less than the saturated model size of 1024 for a 5-mer model.
\texttt{SHazaM} estimated 1015 unique $\theta$ values out of a maximum of 1024 while \texttt{samm} only estimated 137 unique $\theta$ values and logistic estimated 485.
Visually, estimates from the three models look similar, with similar hot- and cold-spots, though \texttt{SHazaM} is more ``spiky'' than \texttt{samm} and logistic (Figure~\ref{fig:shazam_v_samm_v_logistic}).
In terms of model interpretability, \texttt{samm} or logistic regression seem to be preferable to \texttt{SHazaM} as they produce much more parsimonious models.
The logistic model seems to fit a model that is intermediate to \texttt{samm} and \texttt{SHazaM} in terms of parameter support.

Ideally, we would be able to compare the different methods in terms of their observed data likelihood on a test set.
However due to methodological difficulties and incompatibilities of the methods, we were unable to come up with a concrete way to compare the methods.
In particular, \texttt{SHazaM} is not a likelihood-based method.
In addition, the observed data likelihood for \texttt{samm} is computationally intractable, which makes it difficult to compare to other likelihood-based methods.
We hope to come up with a good solution for assessing \texttt{samm} on real-world data in the future.

\section{Discussion}

We have modeled somatic hypermutation of BCR sequences using Cox proportional hazards.
Due to the context-dependence of mutation rates, we must take into account the unknown mutation order to compute the full likelihood.
To deal with this missing data, we used MCEM, where we marginalize over the possible mutation orders using Markov chain Monte Carlo.
Unlike current methods, our regression framework can model the effect of arbitrary features, such as varying motif lengths and sequence positions.
In this paper, we use the lasso to perform feature selection and stabilize our estimates in high-dimensional settings.
One can easily extend this approach to use other sparsity-inducing penalties to reflect other prior beliefs about the model structure.
We show that \texttt{samm} achieves better performance than the state-of-the-art method under a variety of simulation settings.

There are a few limitations with our current method.
We currently subsample our data significantly to ensure our training set is composed of independent observations.
This would not be necessary if we were able to perform accurate phylogenetic ancestral sequence estimation using context-sensitive models.
In addition, our method returns ``uncertainty'' intervals rather than confidence intervals since there are no guarantees on their nominal coverage.
Simulations show that our uncertainty intervals are close to their nominal coverage levels if there is a sufficient amount of data (Figure~\ref{fig:simulation_basic}), but better methods may be available.

While the present analysis only considers sequence context, other biologically-motivated features may be just as informative: nucleotide position, proximity to other contexts, etc.
By incorporating other types of features into the model, we may be able to help verify or find problems with the currently accepted model of somatic hypermutation \citep{Methot2017-gi}.

Finally, our model can be used in other contexts to model other biological processes.
For instance, our method could be used to model the rate of single-nucleotide polymorphisms \citep{Aggarwala2016-bp} and transcription-factor binding \citep{Zhou2004-bz}.

\section*{Acknowledgments}
We are grateful to Duncan Ralph for assistance performing sequence annotation, clustering and simulating germline repertoires.
We would like to thank the Kleinstein lab for generously sharing DNA sequences, and especially to Jason Vander Heiden for providing us with preprocessed versions of their sequence data.

Jean Feng was supported by NIH grants DP5OD019820 and T32CA206089. Noah Simon was supported by NIH grant DP5OD019820.
David Shaw, Vladimir Minin, and Frederick Matsen were supported by NIH grants U19-AI117891 and R01-GM113246; David Shaw and Frederick Matsen were also supported by R01-AI120961.
The research of Frederick Matsen was supported in part by a Faculty Scholar grant from the Howard Hughes Medical Institute and the Simons Foundation.

\bibliographystyle{imsart-nameyear}
\bibliography{motif}

\appendix

\section{Proof of marginal likelihood}

We now prove the statement in Section~\ref{sec:methods} that the marginal likelihood of $\boldsymbol{\theta}$ is given by \eqref{eq:likelihood_complete_data} and only depends on the mutation order $\boldsymbol{\pi}_{1:n}$.

\begin{proof}
	Suppose the mutation times are observed, so $u_{i}$ is the time of the $i$th mutation.
	Then the conditional probability of observing a mutation order $\boldsymbol{\pi}_{1:n}$ given mutation times $\boldsymbol{u} = (u_1, ..., u_n)$ can be written as
	\begin{align}
	\Pr
	\left (
	\boldsymbol{\pi}_{1:n}
	| \boldsymbol{u};
	\boldsymbol{\theta}, h_0
	\right )
	& =
	\prod_{i=1}^n
	\Pr \left (
	\pi_{i}|
	\boldsymbol{\pi}_{1:i - 1},
	u_{i-1}, u_{i} ;
	\boldsymbol{\theta}, h_0
	\right )\\
	& =
	\prod_{i=1}^n
	\frac{
		\Pr\left (
		\pi_{i}, u_{i}|
		\boldsymbol{\pi}_{1:i - 1},
		u_{i-1};
		\boldsymbol{\theta}, h_0
		\right )
	}{
		\sum_{q\in R(\boldsymbol{\pi}_{1:i-1})}
		\Pr \left (
		q, u_{i}|
		\boldsymbol{\pi}_{1:i - 1},
		u_{i-1};
		\boldsymbol{\theta}, h_0
		\right ),
	}
	\label{eq:mut_order_fracs}
	\end{align}
	where the conditional probability of observing a mutation in position $q$ at time $u_i$ is defined as
	\begin{align}
	& \Pr \left (
	q, u_{i}|
	\boldsymbol{\pi}_{1:i - 1},
	u_{i-1};
	\boldsymbol{\theta}, h_0
	\right )
	\label{eq:mut_order_num}
	\\
	& \qquad =
	{h_0}(u_{i})
	\exp \left (
	{\boldsymbol{\theta}}^\top \psi_{q}(S(\boldsymbol{\pi}_{1:i-1}))
	\right )\times
	\label{eq:mut_order_num1}
	\\
	& \qquad \qquad
	\exp
	\left (
	- \sum_{q' \in R(\boldsymbol{\pi}_{1:i-1})}
	\exp \left (
	{\boldsymbol{\theta}}^\top \psi_{q'}(S(\boldsymbol{\pi}_{1:i-1}))
	\right )
	\int_{u_{i-1}}^{u_{i}}
	{h_0}(t)
	dt
	\label{eq:mut_order_num2}
	\right ).
	\end{align}
	Notice that in \eqref{eq:mut_order_num}, the terms ${h_0}(u_{i})$ and \eqref{eq:mut_order_num2} do not depend on $q$.
	So plugging \eqref{eq:mut_order_num} into \eqref{eq:mut_order_fracs}, these two terms cancel and we get
	\begin{align}
	\Pr
	\left (
	\boldsymbol{\pi}_{1:n}
	| \boldsymbol{u};
	\boldsymbol{\theta}, h_0
	\right )
	& =
	\prod_{i=1}^n
	\frac{
		\exp(
		{\boldsymbol{\theta}}^\top \psi_{\pi_{i}}(S(\boldsymbol{\pi}_{1:i-1}))
		)
	}{
		\sum_{q\in R(\boldsymbol{\pi}_{1:i-1})}
		\exp(
		{\boldsymbol{\theta}}^\top \psi_{q}(S(\boldsymbol{\pi}_{1:i-1}))
	)}.
	\label{eq:mut_order_simplified}
	\end{align}
	Since the conditional probability of the mutation order does not depend on mutation times $\boldsymbol{u}$, then the marginal probability of the mutation order
	$\Pr(\pi_{1:n}; \boldsymbol{\theta}, h_0)$
	is also equal to \eqref{eq:mut_order_simplified}.
\end{proof}

\section{Pre-processing data}

If we are interested in modeling the effect of $k$-mer motifs on the hazard rate where $k > 1$, then the positions at the ends of the B-cell receptor sequences must be properly handled.
The issue is that the positions at the ends might not have enough neighboring nucleotides to fully construct a $k$-mer motif.

In order to deal with this issue, we first preprocess our data by trimming the two ends of the BCR sequences until the ends of the na\"ive and mutated sequences are the same.
We then assume that these end positions are fixed and not part of the mutation process.

For example, if we are interested in modeling how 3-mer motifs affect the mutation rate of the center position, we need to handle the special case of the two positions at the ends of the sequence.
Given a na\"ive BCR sequence and its associated mutated sequence, we trim away the positions at the ends of both sequences until the first and last positions are the same.
If our trimmed sequence is of length $p'$, we suppose that only positions 2 through $p' - 1$ can undergo mutation.
We can now apply our estimation method since all positions use the same feature vector mapping.

\section{Other simulations}

\subsection{Reconstructing mutation history}
\label{appendix:parsimony}

Since most clonal families contain multiple sequences, including all sequences without reconstructing the shared mutation history within each family can introduce bias by considering some mutations more than once.
To overcome this bias, we consider two approaches: we can either attempt to estimate this history using standard methods, or we can randomly sample a single sequence from each clonal family.
For the former case, to date, there are no methods that incorporate context-specific mutation models; we introduce one standard and useful approach to consider for a single clonal family.

Assume we have a collection of nucleotide sequences that have mutated away from a known na\"ive sequence.
In time, as we mutate away from this na\"ive sequence, a series of intermediate nucleotide sequences are introduced on the way to obtaining the mutated sequences.
These intermediate sequences, known as ``ancestral states,'' are related to one another and to our observed sequences by an unknown phylogeny: a tree of dependencies that ties all sequences together by common ancestry.
For a comprehensive treatment of phylogenetics, see \citet{felsenstein2003inferring}.

Unfortunately we do not observe these ancestral states.
A simple approach to estimate them is to use parsimony imputation \citep{farris1970methods}, a method that minimizes the total number of mutations that occur on the tree.
Reconstructing ancestral states using parsimony with \texttt{dnapars} \citep{phylip} involves searching through a number of candidate trees and computing the minimum number of changes necessary to obtain each tree.
Among the equally parsimonious trees returned by \texttt{dnapars}, we choose the first one to compute mutation contexts.

To determine the optimal data processing strategy between sampling, imputing ancestral states, and including all sequences without imputation, we simulate 3000 clonal families with realistic sizes.
The composition for each of these clonal families is determined by sampling at random a cluster size and an inferred na\"ive sequence from the \texttt{partis}-processed \citet{Cui2016-wz} dataset.
Cluster sizes range from 1--109.
The median cluster size is two, and about 42\% of all clusters are singletons.
Sequences are 2.5\% mutated on average.
We take $\boldsymbol{\theta}$ to be a random resampling of $\boldsymbol{\theta}_{\MK}$ parameters \citep{Cui2016-wz}.

\begin{table}
\caption{Statistics on reconstructing $\boldsymbol{\theta}$ using various data preprocessing methods.}
    \label{tab:parsimony}
\begin{tabular}{llrr}
\toprule
         &                 &  Relative $\boldsymbol\theta$ error &  Kendall's tau \\
Data processing & Model &                                     &                \\
\midrule
\multirow{2}{*}{All data} & \texttt{SHazaM} &                               0.677 &          0.595 \\
         & \texttt{samm} &                               0.515 &          0.657 \\
\multirow{2}{*}{Imputation} & \texttt{SHazaM} &                               0.721 &          0.580 \\
         & \texttt{samm} &                               0.537 &          0.639 \\
\multirow{2}{*}{Sampling} & \texttt{SHazaM} &                               0.721 &          0.515 \\
         & \texttt{samm} &                               0.500 &          0.647 \\
\bottomrule
\end{tabular}
\end{table}

In Table~\ref{tab:parsimony} we see imputing ancestors using parsimony does not provide any improvements in the model fit in most cases.
Given that mutations in the simulation above occur based on the sequence context, the relatively poor performance of imputing ancestors may be due to the heterogeneity of mutation rates among sites \citep{ho2004-tracing}.
Sampling a random descendant from each clonal family decreases the relative error for \texttt{samm}.
For \texttt{SHazaM}, using all of the data results in the lowest relative error, most likely due to the fact that \texttt{SHazaM} fits mutabilities differently when not enough observations are present, and this case has more data than in the case of sampling.
In Section~\ref{sec:data}, we sample from each clonal family to estimate the fit for \texttt{samm} while using all of the data for \texttt{SHazaM}.

\subsection{Model misspecification: mutating with replacement}
\label{sec:mutating_replacement}
Throughout this manuscript, we have assumed that the positions in a BCR sequence mutate at most once.
This assumption is for computational simplicity: if a position can mutate more than once, our estimation procedure must consider every single possible nucleotide sequence.
However, this may not be realistic biologically.
In this section, we present a simulation study to see how \texttt{samm}'s accuracy changes when positions are allowed to mutate multiple times.

Much of the simulation settings are similar to before.
For the somatic hypermutation model, we resample from $\boldsymbol{\theta}_{\MK}$ -- defined in Section~\ref{sec:sim_examples} -- for each 3-mer motif, then randomly set half of them to zero.
Each dataset consists of 300 simulated BCR sequences from a single mouse.
Mutations are simulated using a survival model where each position can mutate multiple times versus at most one time.
This simulation study is run twenty times.

For low mutation rates of 1--5\%, we have similar accuracy when the model is misspecified (Table~\ref{tab:mut_with_replacement}).
The accuracies are similar since a position is very unlikely to mutate more than once in a low mutation rate setting.

We also try higher mutation rates as it is common to see mutation rates of 5--15\% in humans, especially in individuals with chronic viral infections \citep{He2014-fn}.
Even in this scenario with higher mutation rates, the accuracies are still similar.
These results suggest that our simplifying assumption gives up very little accuracy for a huge gain in computational efficiency.

\begin{table}
	\caption{Results on twenty replicates of simulated data with standard errors (SE).}
	\label{tab:mut_with_replacement}
	\begin{tabular}{rrrr}
		\toprule
		Mutation Rate (\%) & True Model &  Relative $\boldsymbol{\theta}$ error (SE) & Kendall's tau (SE) \\
		\midrule
		\multirow{2}{*}{1--5} & Mutate at most once  & 0.364 (0.015) & 0.722 (0.008)\\
		& Mutate multiple times & 0.348 (0.014) & 0.723 (0.008)\\
		\midrule
		\multirow{2}{*}{5--15} & Mutate at most once  & 0.194 (0.010) & 0.791 (0.008)\\
		& Mutate multiple times & 0.190 (0.010) & 0.781 (0.007)\\
		\bottomrule
	\end{tabular}
\end{table}

\subsection{Simulation results for the 27 settings}
\label{sec:all_the_others}

We report the results from the full set of possible simulation settings from Section~\ref{sec:simulations1} for the unpenalized (Tables~\ref{tab:supp_refit_sim} and~\ref{tab:supp_refit_sim_coverage}) and penalized (Table~\ref{tab:supp_penalized_sim}) fits.
Settings reported in the main manuscript were run 100 times; the others were run ten times.
Across all 2700 simulation runs, a total of twenty replicates fail to obtain confidence intervals after eighty MCEM iterations, given in Table~\ref{tab:failed_runs}.

\begin{table}
	\caption{Number of failed replicates, i.e.\ replicates where variance estimates are negative, out of total number of failed replicates for the simulations}
	\label{tab:failed_runs}
	\begin{tabular}{llll}
		\toprule
		&     &     & Model: failed reps/total reps\\
		\% effect size & \% nonzeros & \# of samples &\\
		\midrule
		50  & 50  & 200 &   2,3-mer: 1/100\\
		100 & 25  & 200 &   2,3-mer: 2/10\\
		& 50  & 100 &   2,3-mer: 1/100\\
		&     &     &   3-mer per-target: 1/100 \\
		&     & 200 &   2,3-mer: 1/100\\
		&     & 400 &   2,3-mer: 7/100\\
		& 100 & 200 &   2,3-mer: 2/100\\
		200 & 50  & 200 &   2,3-mer: 1/100\\
		&     &     &   3-mer per-target: 1/100 \\
		&     & 400 &   3-mer per-target: 1/10 \\
		& 100 & 100 &   3-mer per-target: 1/10 \\
		&     & 200 &   3-mer: 1/10\\
		\bottomrule
	\end{tabular}
\end{table}

In most cases, the penalized fits and unpenalized fits obtain similar relative errors and rank correlations.
In roughly half of cases, the penalized fits obtain smaller relative errors than the unpenalized fits; this may be an effect of the shrinkage present in the penalized $\boldsymbol{\theta}$.
In all cases but four, the unpenalized fits have higher correlation.
We prefer unpenalized fits as they are the only way to obtain reliable uncertainty estimates, though if reconstructing $\boldsymbol{\theta}$ is the primary goal then penalized fits provide a quicker solution.

For the unpenalized fits, the average number of false positives is less than one in the majority of settings, indicating our procedure has good support recovery.
Our model has the most false positives with hierarchical fits on large numbers of samples.
Moreover, we see expected trends in the output -- relative error decreases and correlation increases as sample size increases, and per-target models are more difficult to fit than same-target ones.
Varying effect size and sparsity levels does not seem to affect our method's performance significantly.

\begin{sidewaystable}
\scriptsize
\centering
\caption{
Full simulation results using the unpenalized refit $\boldsymbol{\theta}$.
}
\label{tab:supp_refit_sim}
\begin{tabular}{lll|lll|lll}
\toprule
    &     &     & \multicolumn{3}{c}{Relative $\boldsymbol{\theta}$ error (SE)} &  \multicolumn{3}{c}{Kendall's tau (SE)}\\
    &     &     &              2,3-mer &          3-mer & 3-mer per-target &        2,3-mer &          3-mer & 3-mer per-target \\
\% effect size & \% nonzeros & \# samples &                      &                &                  &                &                &                  \\
\midrule
50  & 25  & 100 &        0.725 (0.156) &  0.836 (0.205) &    1.027 (0.197) &  0.573 (0.061) &  0.504 (0.061) &    0.393 (0.040) \\
    &     & 200 &        0.474 (0.068) &  0.558 (0.098) &    0.846 (0.062) &  0.709 (0.018) &  0.575 (0.063) &    0.477 (0.057) \\
    &     & 400 &        0.380 (0.091) &  0.377 (0.110) &    0.656 (0.052) &  0.777 (0.030) &  0.612 (0.036) &    0.564 (0.035) \\
    & 50  & 100 &        0.682 (0.140) &  0.697 (0.103) &    1.017 (0.106) &  0.625 (0.042) &  0.606 (0.056) &    0.433 (0.033) \\
    &     & 200 &        0.562 (0.196) &  0.572 (0.153) &    0.880 (0.089) &  0.712 (0.044) &  0.670 (0.046) &    0.528 (0.035) \\
    &     & 400 &        0.393 (0.175) &  0.435 (0.081) &    0.679 (0.026) &  0.805 (0.026) &  0.720 (0.041) &    0.647 (0.029) \\
    & 100 & 100 &        0.710 (0.208) &  0.840 (0.194) &    1.217 (0.236) &  0.614 (0.065) &  0.624 (0.023) &    0.458 (0.041) \\
    &     & 200 &        0.547 (0.100) &  0.603 (0.088) &    0.880 (0.097) &  0.708 (0.026) &  0.708 (0.026) &    0.552 (0.045) \\
    &     & 400 &        0.370 (0.047) &  0.468 (0.057) &    0.718 (0.080) &  0.777 (0.028) &  0.769 (0.017) &    0.648 (0.021) \\
100 & 25  & 100 &        0.443 (0.143) &  0.424 (0.119) &    0.680 (0.067) &  0.725 (0.059) &  0.614 (0.047) &    0.535 (0.051) \\
    &     & 200 &        0.331 (0.089) &  0.316 (0.091) &    0.565 (0.053) &  0.777 (0.029) &  0.650 (0.078) &    0.625 (0.036) \\
    &     & 400 &        0.246 (0.046) &  0.245 (0.092) &    0.406 (0.044) &  0.807 (0.021) &  0.649 (0.025) &    0.704 (0.025) \\
    & 50  & 100 &        0.519 (0.148) &  0.539 (0.106) &    0.820 (0.115) &  0.752 (0.033) &  0.695 (0.050) &    0.567 (0.039) \\
    &     & 200 &        0.355 (0.090) &  0.369 (0.073) &    0.658 (0.051) &  0.812 (0.028) &  0.745 (0.039) &    0.658 (0.035) \\
    &     & 400 &        0.248 (0.049) &  0.275 (0.060) &    0.496 (0.046) &  0.860 (0.020) &  0.779 (0.032) &    0.743 (0.026) \\
    & 100 & 100 &        0.496 (0.063) &  0.631 (0.101) &    0.852 (0.062) &  0.712 (0.033) &  0.742 (0.027) &    0.568 (0.034) \\
    &     & 200 &        0.373 (0.057) &  0.429 (0.062) &    0.649 (0.061) &  0.783 (0.026) &  0.805 (0.031) &    0.692 (0.026) \\
    &     & 400 &        0.270 (0.031) &  0.318 (0.064) &    0.483 (0.035) &  0.842 (0.017) &  0.864 (0.020) &    0.770 (0.017) \\
200 & 25  & 100 &        0.483 (0.124) &  0.499 (0.069) &    0.639 (0.050) &  0.689 (0.025) &  0.610 (0.030) &    0.559 (0.049) \\
    &     & 200 &        0.357 (0.047) &  0.389 (0.046) &    0.534 (0.034) &  0.737 (0.043) &  0.637 (0.037) &    0.619 (0.042) \\
    &     & 400 &        0.272 (0.033) &  0.304 (0.024) &    0.467 (0.035) &  0.792 (0.024) &  0.708 (0.055) &    0.690 (0.022) \\
    & 50  & 100 &        0.420 (0.065) &  0.494 (0.044) &    0.633 (0.033) &  0.755 (0.029) &  0.648 (0.044) &    0.594 (0.029) \\
    &     & 200 &        0.317 (0.042) &  0.411 (0.046) &    0.555 (0.036) &  0.811 (0.025) &  0.695 (0.034) &    0.667 (0.028) \\
    &     & 400 &        0.257 (0.028) &  0.316 (0.021) &    0.501 (0.043) &  0.857 (0.017) &  0.760 (0.029) &    0.714 (0.029) \\
    & 100 & 100 &        0.459 (0.039) &  0.532 (0.056) &    0.684 (0.081) &  0.680 (0.038) &  0.747 (0.025) &    0.618 (0.024) \\
    &     & 200 &        0.378 (0.053) &  0.388 (0.035) &    0.533 (0.041) &  0.741 (0.040) &  0.808 (0.020) &    0.716 (0.017) \\
    &     & 400 &        0.312 (0.029) &  0.323 (0.033) &    0.454 (0.030) &  0.800 (0.033) &  0.852 (0.016) &    0.771 (0.013) \\
\bottomrule
\end{tabular}
\end{sidewaystable}

\begin{sidewaystable}
\scriptsize
\centering
\caption{
Full simulation results reporting coverage and false discovery statistics using the unpenalized refit $\boldsymbol{\theta}$.
}
\label{tab:supp_refit_sim_coverage}
\begin{tabular}{lll|lll|lll}
\toprule
    &     &     & \multicolumn{3}{c}{Coverage (SE)} &  \multicolumn{3}{c}{Num False Positive; Num Discovered (SE; SE)}\\
    &     &     &              2,3-mer &          3-mer & 3-mer per-target &        2,3-mer &          3-mer & 3-mer per-target \\
\% effect size & \% nonzeros & \# samples &                      &                &                  &                &                &                  \\
\midrule
50  & 25  & 100 &  85.4 (10.5) &  89.7 (10.2) &      72.2 (11.9) &                1.2; 6.9 (0.7; 1.5) &    1.5; 8.9 (0.8; 2.0) &   2.0; 13.0 (1.6; 4.5) \\
    &     & 200 &   93.8 (5.6) &   87.9 (8.8) &      67.4 (13.9) &                0.9; 9.0 (0.8; 2.9) &   2.5; 12.7 (1.7; 1.6) &   4.9; 24.0 (1.7; 2.6) \\
    &     & 400 &   96.4 (1.7) &   93.6 (4.8) &      81.6 (10.6) &               1.4; 13.2 (0.8; 2.0) &   2.5; 15.0 (1.8; 1.7) &   4.0; 32.7 (2.7; 4.3) \\
    & 50  & 100 &   85.8 (9.5) &   93.7 (3.6) &      71.5 (14.5) &                0.9; 8.7 (0.8; 2.7) &   1.3; 15.1 (1.0; 2.8) &   0.3; 15.7 (0.5; 4.7) \\
    &     & 200 &   94.4 (6.0) &   94.7 (3.5) &      75.2 (12.8) &               1.6; 11.1 (1.4; 4.0) &   1.1; 18.3 (1.0; 2.1) &   1.6; 26.1 (1.3; 4.8) \\
    &     & 400 &   98.9 (1.6) &   93.6 (4.2) &       87.3 (7.9) &               2.2; 13.5 (2.6; 6.7) &   1.5; 23.8 (1.3; 2.0) &   1.6; 43.6 (1.2; 4.8) \\
    & 100 & 100 &   89.7 (9.7) &   92.7 (4.0) &       80.0 (9.8) &                0.0; 7.1 (0.0; 1.9) &   0.0; 18.0 (0.0; 1.7) &   0.0; 23.3 (0.0; 5.3) \\
    &     & 200 &   96.1 (3.7) &   95.7 (1.9) &       81.1 (9.3) &                0.0; 6.3 (0.0; 2.5) &   0.0; 24.0 (0.0; 2.9) &   0.0; 32.5 (0.0; 4.9) \\
    &     & 400 &   96.7 (3.1) &   96.2 (3.2) &       92.2 (6.2) &               0.0; 10.5 (0.0; 4.3) &   0.0; 30.3 (0.0; 2.7) &   0.0; 55.3 (0.0; 6.1) \\
100 & 25  & 100 &   96.7 (2.3) &   92.2 (9.3) &      76.1 (11.2) &                0.3; 8.5 (0.6; 3.1) &   1.5; 10.6 (1.3; 2.0) &   2.3; 21.2 (1.5; 4.4) \\
    &     & 200 &   95.3 (4.4) &   92.5 (5.2) &       84.2 (9.6) &               1.1; 12.5 (1.4; 4.0) &   1.8; 13.3 (1.3; 2.3) &   2.5; 30.3 (1.9; 5.4) \\
    &     & 400 &   97.4 (1.7) &   93.4 (4.6) &       87.5 (9.8) &               1.0; 14.6 (0.7; 2.6) &   2.2; 15.5 (1.8; 2.1) &   4.9; 46.7 (2.0; 4.2) \\
    & 50  & 100 &   96.7 (4.4) &   96.0 (3.7) &       79.0 (9.4) &               1.3; 10.9 (1.3; 4.7) &   0.8; 16.7 (0.9; 2.6) &   1.1; 27.9 (1.3; 6.2) \\
    &     & 200 &   97.3 (3.3) &   96.0 (4.0) &       85.0 (9.8) &               2.3; 16.0 (2.1; 6.2) &   0.9; 21.5 (0.9; 2.5) &   0.9; 40.0 (0.9; 5.8) \\
    &     & 400 &   97.7 (2.5) &   95.3 (4.1) &       89.8 (7.6) &               4.5; 21.2 (2.5; 7.2) &   1.1; 24.2 (1.0; 4.6) &   1.9; 59.1 (1.3; 6.9) \\
    & 100 & 100 &   95.6 (6.7) &   96.0 (4.1) &       84.0 (8.7) &                0.0; 6.7 (0.0; 5.1) &   0.0; 22.3 (0.0; 6.4) &   0.0; 39.4 (0.0; 5.3) \\
    &     & 200 &   95.7 (4.1) &   96.2 (4.4) &       91.7 (6.5) &               0.0; 13.0 (0.0; 6.2) &   0.0; 31.6 (0.0; 8.1) &   0.0; 63.1 (0.0; 8.0) \\
    &     & 400 &   95.8 (2.7) &   97.1 (3.3) &       94.6 (3.4) &               0.0; 24.8 (0.0; 3.8) &  0.0; 34.2 (0.0; 15.5) &  0.0; 86.8 (0.0; 14.3) \\
200 & 25  & 100 &   99.5 (1.0) &   93.4 (2.2) &       83.9 (8.0) &                0.2; 4.7 (0.4; 2.4) &   2.1; 10.4 (1.0; 1.7) &   1.3; 23.7 (1.1; 2.3) \\
    &     & 200 &   98.4 (2.1) &   88.5 (6.6) &       83.5 (5.6) &                0.8; 9.8 (1.0; 4.0) &   2.3; 12.6 (1.6; 1.8) &   1.5; 29.7 (1.0; 2.6) \\
    &     & 400 &   95.9 (6.0) &   88.9 (7.0) &       87.4 (6.5) &               3.7; 16.3 (3.0; 6.1) &   1.7; 14.4 (1.6; 1.9) &   2.2; 36.0 (1.9; 5.2) \\
    & 50  & 100 &   99.2 (1.1) &   96.5 (2.8) &       92.3 (4.8) &               0.5; 10.7 (0.9; 3.5) &   0.7; 15.0 (0.8; 1.7) &   0.3; 30.3 (0.6; 6.3) \\
    &     & 200 &   98.2 (2.5) &   95.7 (5.9) &       93.1 (5.2) &               3.1; 17.5 (2.1; 6.1) &   0.6; 14.9 (0.9; 3.9) &   0.2; 39.6 (0.6; 6.7) \\
    &     & 400 &   95.0 (5.3) &   95.6 (5.5) &       89.9 (9.2) &               6.7; 25.8 (2.9; 5.7) &   0.2; 18.1 (0.4; 3.1) &   0.6; 48.6 (0.7; 9.9) \\
    & 100 & 100 &   98.0 (2.0) &   98.5 (3.4) &       91.3 (4.1) &                0.0; 8.2 (0.0; 4.2) &   0.0; 16.0 (0.0; 7.8) &   0.0; 46.2 (0.0; 2.9) \\
    &     & 200 &   96.4 (7.8) &   93.1 (7.1) &       95.4 (4.7) &               0.0; 11.9 (0.0; 5.5) &   0.0; 24.8 (0.0; 6.6) &  0.0; 56.4 (0.0; 12.4) \\
    &     & 400 &   93.6 (7.5) &   98.5 (2.2) &       96.6 (1.9) &               0.0; 25.1 (0.0; 6.4) &  0.0; 28.9 (0.0; 12.3) &  0.0; 80.3 (0.0; 12.1) \\
\bottomrule
\end{tabular}
\end{sidewaystable}

\begin{sidewaystable}
\scriptsize
\centering
\caption{
Full simulation results using penalized $\boldsymbol{\theta}$.
}
\label{tab:supp_penalized_sim}
\begin{tabular}{lll|lll|lll}
\toprule
    &     &     & \multicolumn{3}{c}{Relative $\boldsymbol{\theta}$ error (SE)} &  \multicolumn{3}{c}{Kendall's tau (SE)}\\
    &     &     &              2,3-mer &          3-mer & 3-mer per-target &        2,3-mer &          3-mer & 3-mer per-target \\
\% effect size & \% nonzeros & \# samples &                      &                &                  &                &                &                  \\
\midrule
50  & 25  & 100 &        0.539 (0.064) &  0.671 (0.083) &    0.809 (0.058) &  0.555 (0.059) &  0.503 (0.063) &    0.396 (0.036) \\
    &     & 200 &        0.438 (0.052) &  0.521 (0.069) &    0.698 (0.060) &  0.660 (0.049) &  0.561 (0.060) &    0.454 (0.059) \\
    &     & 400 &        0.332 (0.041) &  0.391 (0.086) &    0.572 (0.041) &  0.754 (0.036) &  0.586 (0.036) &    0.543 (0.036) \\
    & 50  & 100 &        0.558 (0.056) &  0.553 (0.037) &    0.780 (0.041) &  0.608 (0.046) &  0.595 (0.062) &    0.427 (0.031) \\
    &     & 200 &        0.464 (0.090) &  0.508 (0.100) &    0.701 (0.028) &  0.702 (0.048) &  0.660 (0.046) &    0.523 (0.030) \\
    &     & 400 &        0.361 (0.079) &  0.393 (0.057) &    0.599 (0.030) &  0.788 (0.025) &  0.711 (0.042) &    0.615 (0.034) \\
    & 100 & 100 &        0.555 (0.066) &  0.692 (0.087) &    0.824 (0.048) &  0.617 (0.055) &  0.608 (0.031) &    0.460 (0.050) \\
    &     & 200 &        0.466 (0.031) &  0.562 (0.034) &    0.722 (0.044) &  0.700 (0.022) &  0.681 (0.039) &    0.546 (0.031) \\
    &     & 400 &        0.368 (0.027) &  0.451 (0.048) &    0.622 (0.050) &  0.770 (0.019) &  0.744 (0.027) &    0.631 (0.030) \\
100 & 25  & 100 &        0.399 (0.112) &  0.425 (0.092) &    0.706 (0.063) &  0.690 (0.063) &  0.608 (0.053) &    0.502 (0.047) \\
    &     & 200 &        0.305 (0.067) &  0.383 (0.090) &    0.550 (0.066) &  0.757 (0.038) &  0.646 (0.076) &    0.600 (0.038) \\
    &     & 400 &        0.211 (0.031) &  0.296 (0.078) &    0.439 (0.034) &  0.800 (0.026) &  0.643 (0.026) &    0.678 (0.024) \\
    & 50  & 100 &        0.413 (0.102) &  0.465 (0.055) &    0.685 (0.038) &  0.740 (0.040) &  0.681 (0.045) &    0.548 (0.040) \\
    &     & 200 &        0.315 (0.060) &  0.360 (0.067) &    0.592 (0.043) &  0.805 (0.031) &  0.735 (0.039) &    0.633 (0.038) \\
    &     & 400 &        0.240 (0.036) &  0.281 (0.058) &    0.503 (0.044) &  0.848 (0.023) &  0.768 (0.032) &    0.713 (0.034) \\
    & 100 & 100 &        0.430 (0.038) &  0.540 (0.033) &    0.696 (0.051) &  0.712 (0.041) &  0.724 (0.027) &    0.579 (0.036) \\
    &     & 200 &        0.362 (0.052) &  0.421 (0.053) &    0.580 (0.035) &  0.763 (0.035) &  0.783 (0.032) &    0.670 (0.028) \\
    &     & 400 &        0.272 (0.025) &  0.332 (0.038) &    0.490 (0.041) &  0.826 (0.008) &  0.839 (0.023) &    0.742 (0.025) \\
200 & 25  & 100 &        0.448 (0.102) &  0.518 (0.018) &    0.674 (0.035) &  0.637 (0.034) &  0.593 (0.044) &    0.495 (0.069) \\
    &     & 200 &        0.359 (0.036) &  0.487 (0.040) &    0.642 (0.017) &  0.701 (0.052) &  0.611 (0.050) &    0.567 (0.057) \\
    &     & 400 &        0.290 (0.032) &  0.453 (0.040) &    0.567 (0.046) &  0.758 (0.039) &  0.707 (0.049) &    0.644 (0.031) \\
    & 50  & 100 &        0.375 (0.025) &  0.462 (0.034) &    0.649 (0.034) &  0.747 (0.023) &  0.633 (0.041) &    0.540 (0.040) \\
    &     & 200 &        0.315 (0.037) &  0.422 (0.038) &    0.590 (0.035) &  0.796 (0.031) &  0.678 (0.034) &    0.624 (0.036) \\
    &     & 400 &        0.279 (0.032) &  0.368 (0.027) &    0.545 (0.043) &  0.838 (0.016) &  0.737 (0.035) &    0.684 (0.031) \\
    & 100 & 100 &        0.466 (0.022) &  0.518 (0.029) &    0.647 (0.033) &  0.652 (0.040) &  0.717 (0.017) &    0.597 (0.024) \\
    &     & 200 &        0.403 (0.049) &  0.433 (0.049) &    0.567 (0.024) &  0.709 (0.042) &  0.777 (0.024) &    0.678 (0.020) \\
    &     & 400 &        0.348 (0.035) &  0.364 (0.044) &    0.503 (0.027) &  0.777 (0.034) &  0.826 (0.026) &    0.735 (0.015) \\
\bottomrule
\end{tabular}
\end{sidewaystable}

\clearpage
\section{Computing the survival process likelihood on a tree with ancestral sequences at internal nodes}
\label{sec:samm_rank}

Sequences evolve along a tree with a shared mutation history.
Often, given a set of sequence data, many candidate trees optimize the maximum parsimony objective function \citep{farris1970methods}, and thus phylogenetic algorithms can return multiple solutions.
We have found it to be useful to rank a set of equally-parsimonious phylogenetic trees in terms of an additional objective function \citep{Davidsen2018-lz,DeWitt2018-el}.
To do such ranking with a motif mutability model requires taking into account mutation order -- though the na\"ive and mutated sequences are the same across trees, the pairs of parent and descendant sequences on each branch are not.
Though tempting, we cannot use the surrogate function \eqref{eq:surrogate} on different trees separately and compare them as the observed data differs between different trees.
Instead we are interested in
\begin{equation}
\log \mathcal{L}(\mathbf{S}_{\text{obs}} ; {\boldsymbol{\theta}}) -
\log \mathcal{L}(\mathbf{S}_{\text{obs}}' ; {\boldsymbol{\theta}} )
\end{equation}
which requires estimating the observed likelihood.

To obtain the observed likelihood of data given a tree with inferred ancestral sequences at internal nodes, we use Chib's method to integrate out mutation order along each branch \citep{chibs1995marginal}.
This gives us an estimate of the observed likelihood and allows us to compare multiple trees fit to the same data.

\end{document}